\documentclass[11pt,a4paper]{amsart}

\usepackage{fullpage,amsthm,amssymb,amsmath,epic,eepic,float,graphicx}

\usepackage{color}

\numberwithin{equation}{section}

\def\ll{\left\lgroup}
\def\rr{\right\rgroup}

\def\leq{\leqslant}

\newcommand{\cC}{\mathcal C}

\newcommand{\cE}{\mathcal E}
\newcommand{\cF}{\mathcal F}
\newcommand{\cH}{\mathcal H}
\newcommand{\cM}{\mathcal M}
\newcommand{\cN}{\mathcal N}
\newcommand{\cO}{\mathcal O}
\newcommand{\cP}{\mathcal P}

\newcommand{\bX}{\bar{X}}
\newcommand{\bY}{\bar{Y}}
\newcommand{\bZ}{\bar{Z}}

\def\tr{\operatorname{tr}}

\def\det{\operatorname{det}}

\newcommand{\be}{\begin{eqnarray}}
\newcommand{\ee}{\end{eqnarray}}
\newcommand{\non}{\nonumber}

\newcommand{\LL}{\ensuremath{L}}

\begin{document}
\strut\hfill NORDITA-2012-12, UMTG--273
\vspace{.5in}
\title[]
{OPE in planar QCD from integrability}
\author{
Changrim Ahn $^1$,
         Omar Foda $^2$ and
         Rafael I Nepomechie $^3$
}
\address{
\!\!\!\!\!\!\!$^1$ Department of Physics,
     Ewha Womans University,
     Seoul 120-750, South Korea.  
     \newline
$^2$ Dept of Mathematics and Statistics,
     University of Melbourne,
     Victoria 3010, Australia.  
     \newline
$^3$ Physics Department, P O Box 248046,
     University of Miami,
     Coral Gables, FL 33124 USA.
}
\email{ahn@ewha.ac.kr,
       omar.foda@unimelb.edu.au,
       nepomechie@physics.miami.edu
}
\begin{abstract}
We consider the operator product expansion of local gauge-invariant 
single-trace operators composed of self-dual components of the field 
strength tensor in planar QCD. Using the integrability of the 1-loop
dilatation operator, we obtain a determinant expression for certain 
tree-level structure constants.
\end{abstract}
\maketitle

\setcounter{footnote}{0}

\section{Introduction}
\label{sec:intro}

\subsection{SYM$_4$ and integrable spin chains} 
The problem of computing the conformal dimensions of local, 
gauge-invariant single-trace composite operators in planar 
$\cN \! = \! 4$ supersymmetric Yang-Mills theory 
in (3+1)-dimensions, SYM$_4$, is integrable 
\cite{Minahan:2002ve, Beisert:2003yb, Beisert:2005fw}. 
At 1-loop level, the mixing matrix $\Gamma$ maps 
to the Hamiltonian $\cH_{PSU(2,2 | 4)}$ of an integrable 
$PSU(2,2 | 4)$-symmetric spin chain with nearest-neighbor 
interactions and periodic boundary conditions, such that  
{\bf 1.} The eigenstates $\{\cO\}$ of $\Gamma$ are in 
one-to-one correspondence with the eigenstates 
$\{ | \cO \rangle \}$ of $\cH_{PSU(2,2 | 4)}$, and 
{\bf 2.} The eigenvalues $\{\gamma \}$ of $\Gamma$, 
which are the anomalous dimensions of $\{\cO\}$, 
are equal to the eigenvalues $\{\cE \}$ of 
$\cH_{PSU(2,2 | 4)}$ 
\footnote{\ In this note, $\cO$ is a local gauge-invariant 
single-trace composite operator, in SYM$_4$ or in QCD depending 
on context, that is an eigenstate of the mixing matrix 
$\Gamma$, with anomalous dimension $\gamma$. 
For brevity, we will refer to $\cO$ from now on simply 
as {\it \rq a single-trace operator\rq}.
$| \cO \rangle$ is the corresponding eigenstate of 
the integrable spin chain Hamiltonian $\cH$, whose 
eigenvalue $\cE$ $=$ $\gamma$. The notation $\{\cO\}$ 
stands for sets of single-trace operators,
{\it etc.}}.
Since the eigenstates and eigenvalues of 
$\cH_{PSU(2,2 | 4)}$ can be computed using Bethe ansatz methods, 
the problem is integrable 
\footnote{\ The situation at higher loops is more complicated: 
Spin chains with nearest-neighbor interaction are replaced 
with spin chains with long range interactions, the algebraic 
Bethe ansatz is replaced with an asymptotic Bethe ansatz, and 
finite-size effects must be accounted for. In this note, we 
restrict our attention to 1-loop level and nearest-neighbor 
interacting spin chains.}. 
For a recent review, see \cite{Beisert:2010jr} and references 
therein. 

\subsection{SYM$_4$ $SU(2)$-doublets and spin-$\frac{1}{2}$ 
chains} 
SYM$_4$ contains a vector gauge field, four 
chiral and four anti-chiral spin-$\frac{1}{2}$ fermions, 
and six real scalars that can be expressed as three 
complex scalars 
$\{  X,   Y,   Z\}$ and their charge-conjugates 
$\{\bX, \bY, \bZ\}$.
Any two complex scalars that are not charge
conjugates, such as $\{X, Z\}$ or $\{X, \bZ\}$, mix only 
amongst themselves to form an $SU(2)$-doublet and 
an $SU(2)$-invariant scalar subsector of SYM$_4$. 
In the planar limit at 1-loop level,
the single-trace operators $\{\cO\}$, that are 
composed of a single $SU(2)$ doublet, and that are eigenstates 
of $\Gamma$, map to eigenstates $\{| \cO \rangle\}$ of 
the Hamiltonian $\cH_{\frac{1}{2}}$ of a periodic XXX 
spin-$\frac{1}{2}$ chain.

\subsection{QCD $SU(2)$-triplets and spin-1 chains} 
In \cite{Ferretti:2004ba}, Ferretti, Heise and Zarembo noted that,
at 1-loop level, operators composed of self-dual components 
$\{f_{+}, f_{0}, f_{-}\}$ of 
the QCD field strength tensor mix only among themselves to 
form an $SU(2)$-triplet. Using 
that observation, as well as the fact that QCD with no matter 
fields is conformally invariant (the beta function vanishes) 
in the planar limit at 1-loop level, they showed that local 
single-trace operators 
$\{\cO\}$ that are eigenstates of $\Gamma$ correspond to 
eigenstates of the Hamiltonian $\cH_1$ of an integrable XXX 
spin-1 \cite{Zamolodchikov:1980ku} 
chain. \footnote{\ All spin chains mentioned in this note will be 
integrable (their $R$-matrices satisfy Yang-Baxter equations),
of XXX type (their $R$-matrices are parametrized by rational 
functions in the rapidity variables), and satisfy periodic 
boundary conditions, hence we need not repeat this from now
on.}
As in the spin-$\frac{1}{2}$ case,
the spin-1 chain eigenstates 
and eigenvalues can be computed using Bethe ansatz methods
\cite{Kulish:1981gi, Kulish:1981bi, A.:1982zz, Babujian:1983ae}.  

\subsection{SYM$_4$ structure constants}
Following 
\cite{Lee:1998bxa, Okuyama:2004bd, Roiban:2004va, Alday:2005nd}, 
Escobedo, Gromov, Sever and Vieira 
\cite{Escobedo:2010xs} 
used the connection to spin-$\frac{1}{2}$ chains to obtain a sum 
expression for the structure constants of 3-point functions of 
single-trace operators $\{\cO\}$ in $SU(2)$ 
scalar subsectors of SYM$_4$. They noted that the 
three operators 
$\cO_i$, of lengths $L_i$, $i \in \{1, 2, 3\}$, could be chosen 
to be non-BPS (their conformal dimensions are unprotected by 
supersymmetry) and non-extremal ($L_i < L_j + L_k$, for any 
choice of distinct $i$, $j$ and $k$).

In \cite{Foda:2011rr}, the sum expression of Escobedo {\it et al.}
was evaluated in determinant form. This was made possible by 
the fact that, when expressed in spin chain terms, the essential 
factor in the sum expression can be identified with (a special 
case of) the scalar product of an eigenstate of $\cH_{\frac{1}{2}}$ 
and a generic state (not an eigenstate of $\cH_{\frac{1}{2}}$).

\subsection{QCD structure constants}
In this note, we extend the results of 
\cite{Escobedo:2010xs, Foda:2011rr}, 
from SYM$_4$ and spin-$\frac{1}{2}$ chains to QCD and spin-1 chains, 
to gain information about QCD operator product expansions, OPE's, 
of the operators $\{\cO\}$ of Ferretti {\it et al.}  
\footnote{\ Our results are subject to the same restrictions as 
in \cite{Ferretti:2004ba}, and are valid only in
the planar limit ($N_{c} \rightarrow \infty$ 
and $g \rightarrow 0$, with $\lambda=g^{2} N_{c}$ constant)
and at one-loop level,  
so that the beta function vanishes, and the theory is conformally 
invariant.}. 

We show that 
{\bf 1.} In the general case where all three operators 
$\cO_i$, $i \in \{1, 2, 3\}$ are non-BPS-like (all three states 
map to eigenstates of $\cH_1$ that are not spin-chain reference 
states), the tree-level structure constants can be expressed in 
a sum form that is similar to, but even less restricted than that 
of Escobedo {\it et al.} 
\footnote{\ The sum form of Escobedo {\it et al.} involves 
a summation over all partitions of one set of rapidity 
variables. The sum form that we obtain in the general 
case of three non-BPS-like operators involves 
summations over all partitions of three sets of rapidity 
variables with constraints between them.}
{\bf 2.} In the special case where one operator, {\it e.g.} 
$\cO_3$, is BPS-like  (it maps to a spin-chain reference state), 
the tree-level structure constants can be expressed in a determinant 
form that is similar to that in \cite{Foda:2011rr}. 

In other words, to express the tree-level structure constants in 
determinant form, (at least) one of the three operators must be 
BPS-like. In the following subsection, we outline why this is 
the case. More details are given in Section 
{\bf \ref{sec:structureconstants}}.

\subsection{SYM$_4$ structure constants that can be evaluated 
as determinants} 
The SYM$_4$ structure constants studied in 
\cite{Escobedo:2010xs, Foda:2011rr} 
involve four types of scalars, $\{X, Z, \bar X, \bar Z\}$. 
The only non-vanishing Wick contractions (2-point functions) 
are those between charge-conjugate pairs, that is 
$\langle   X \bX \rangle$, 
$\langle \bX   X \rangle$, 
$\langle   Z \bZ \rangle$, or
$\langle \bZ   Z \rangle$. 
Each operator $\cO_i$, $i \in \{1, 2, 3\}$, consists  
of two types of non-conjugate scalars, that is 
$\{X,     Z\}$, 
$\{X,   \bZ\}$, 
$\{\bX,   Z\}$, and 
$\{\bX, \bZ\}$. 

If 
$\cO_1$ is $\{ X, Z\}$-type 
(a composite operator of scalars of type $\{ X, Z\}$), and 
$\cO_2$ is $\{\bX, \bZ\}$-type, 
there are non-zero Wick contractions of both types, 
$\langle X \bX \rangle$ and 
$\langle Z \bZ \rangle$, between $\cO_1$ and $\cO_2$. 
Now consider $\cO_3$. There is no way to choose the 
scalar content of $\cO_3$ such that 
{\bf 1.} It has non-zero Wick contractions of both types with $\cO_1$, 
{\bf 2.} It has non-zero Wick contractions of both types with $\cO_2$, 
and
{\bf 3.} The 3-point function is non-extremal, which requires 
that $\cO_3$ has non-zero Wick contractions with both $\cO_1$ 
and $\cO_2$. 
The only way to have a non-extremal 3-point function is 
to choose $\cO_3$ to be 
$\{\bX,   Z\}$-type or 
$\{  X, \bZ\}$-type.
Either way, the Wick contractions between 
$\cO_1$ and $\cO_3$ will be of one type only,
and the Wick contractions between 
$\cO_2$ and $\cO_3$ will also be of one type only,
different from that between $\cO_1$ and $\cO_2$. 
These constraints simplify the structure constant 
and allow one to evaluate the sum form 
of Escobedo {\it et al.} in determinant form.

\subsection{QCD structure constants that can be evaluated 
as determinants} 
The QCD structure constants studied in this note 
involve three types of scalars, $\{f_{+}, f_{0}, f_{-}\}$. 
The non-vanishing Wick contractions are those between 
spin-conjugate pairs, that is 
$\langle f_{+} f_{-} \rangle$, 
$\langle f_{-} f_{+} \rangle$, 
and 
$\langle f_{0} f_{0} \rangle$.

Since the action of the Bethe creation operators on the spin-1
reference states generates all three scalars, each operator 
$\cO_i$, $i \in \{1, 2, 3\}$, will consist of all three scalars.
Consequently, there are no constraints on the Wick contractions, 
and the 3-point function of non-BPS operators is more complicated 
than in the SYM$_4$ case  
\footnote{\ In particular, while integrable spin-1 chains 
are related to integrable spin-$\frac{1}{2}$ chains by fusion, 
there is no way that one can use fusion to obtain a 3-point 
function of non-BPS-like operators in the spin-1 case from the 
corresponding spin-$\frac{1}{2}$ result.}.
This 3-point function between three non-BPS-like operators can 
be expressed in sum form, as we will explain in the sequel, but 
that sum form will be more complicated than that 
in \cite{Escobedo:2010xs}, and less useful.

The aim of this note is to identify the structure constants 
that can be evaluated in single determinant form using currently 
available methods of integrability 
\footnote{\ What we have in mind is Slavnov's determinant 
expression for the scalar product of an eigenstate of 
the Hamiltonian and a generic state. This determinant 
expression is unique. It is conceivable that 
determinant expressions for more general scalar products, 
that will allow us to evaluate more general structure 
constants, will eventually be found, but this is obviously 
beyond the scope of this work.}.
Our result is that, in QCD and the spin-1 case, determinant 
expressions for the structure constants require that one 
operator is BPS-like. In other words, that it maps to a
spin chain reference state. 

\subsection{Outline of contents} 
In Section {\bf \ref{sec:ops}}, 
we review the construction of the single-trace 
composite operators from the self-dual components of the field 
strength tensor, the 1-loop mixing matrix, operator product 
expansions, and the {\it \lq tailoring\rq} approach of Escobedo 
{\it et al.} to the structure constants. 
In Section {\bf \ref{sec:BA}}, 
we recall the algebraic Bethe ansatz solution for the eigenstates 
and eigenvalues of the mixing matrix. 
In Section {\bf \ref{sec:structureconstants}}, 
we present our results for the structure constants in terms of 
solutions of the Bethe equations. Section {\bf \ref{sec:discussion}} 
contains a brief discussion. 
In Appendix {\bf \ref{sec:CBA}}, we recall the coordinate Bethe 
ansatz and the $\cF$-conjugation of \cite{Escobedo:2010xs}.
In Appendix {\bf \ref{sec:scalarproducts}}, we present the 
scalar products that appear in the expression for the structure 
constants.

\section{Composite operators, operator product 
expansions and structure constants}\label{sec:ops}

\subsection{Self-dual field-strength components as an $SU(2)$-triplet}
Following \cite{Ferretti:2004ba}, we decompose the QCD Yang-Mills 
field strength tensor 
$F_{\mu\nu}=
\partial_{\mu} A_{\nu} -
\partial_{\nu} A_{\mu} + i g 
\left[A_{\mu} \,, A_{\nu}\right]$ 
into self-dual, $f_{\alpha\beta}$, and 
anti-self-dual, $\bar{f}_{\dot\alpha\dot\beta}$, components,

\be
F_{\mu\nu} = \sigma_{\mu\nu}^{\ \ \alpha\beta}f_{\alpha\beta}+
\bar{\sigma}_{\mu\nu}^{\ \ \dot\alpha\dot\beta} 
\bar{f}_{\dot\alpha\dot\beta}\,,
\ee

\noindent where

\begin{multline}
\sigma_{\mu \nu} =
\frac{i}{4}
\sigma_{2}
\left(\sigma_{\mu}\bar{\sigma}_{\nu}
-
\sigma_{\nu}\bar{\sigma}_{\mu}\right), 
\ 
\bar{\sigma}_{\mu\nu} = 
- \frac{i}{4} \left(\bar{\sigma}_{\mu}\sigma_{\nu}
-\bar{\sigma}_{\nu}\sigma_{\mu}\right)\sigma_{2}, 
\ 
    {\sigma}_{\mu} = (1,  \vec \sigma), 
\  
\bar{\sigma}_{\mu} = (1, -\vec \sigma) \,.
\end{multline}

\noindent We further define
\be
f_{A} = 
\left(
\sigma_{2} \sigma_{A}
\right)^{\alpha \beta} 
f_{\alpha\beta}\,, \qquad 
\bar{f}_{\dot A} = 
\left(
\sigma_{\dot A}\sigma_{2} 
\right)^{\dot \alpha \dot \beta}
\bar{f}_{\dot\alpha\dot\beta}\,,
\ee

\noindent where $A, \dot A = 1,2,3$.
The 2-point function of the field
strength tensor has the structure

\be
\label{FF2ptfunction}
\langle 
F_{\mu \nu\    b}^{\ \ a}(x) 
F_{\rho\sigma\ d}^{\ \ c}(0) 
\rangle = 
\phi(x)
\left( \eta_{\mu\rho} \eta_{\nu\sigma} - \eta_{\mu\sigma} 
\eta_{\nu\rho} \right)\delta^{a}_{d}\delta^{c}_{b}\,,
\ee 

\noindent where $a,b,c,d = 1, \ldots, N_{c}$ are color indices, 
and $\phi(x)$ is a scalar function.  Hence,

\be
\langle f_{A\ b}^{\ a}(x) f_{B\ d}^{\  c}(0) \rangle
= 
\phi(x) \delta_{AB}\delta^{a}_{d}\delta^{c}_{b} 
\,, \qquad
\langle 
f_{A\ b}^{\ a}(x) \bar{f}_{\dot B\ d}^{\  c}(0) 
\rangle = 0 \,.
\label{f2pt}
\ee 

\noindent Following \cite{Beisert:2004fv}, we write

\begin{multline}
f_{+} = f_{11} =\frac{1}{2} \left( f_{2} + i f_{1} \right), 
\ \  
f_{0} = \frac{1}{\sqrt{2}} \left( f_{12} + f_{21} \right) = 
-\frac{i}{\sqrt{2}} f_{3}, 
\ \  
f_{-} = f_{22} =\frac{1}{2} \left( f_{2} - i f_{1} \right) \,.
\end{multline}

\noindent From Equation ({\bf \ref{f2pt}}),
$\langle f_{\pm}(x) f_{\pm}(0) \rangle =
 \langle f_{\pm}(x) f_{0}  (0) \rangle = 0$, 
and the only nonzero Wick contractions (2-point functions) 
are 
$\langle f_{\pm} f_{\mp}\rangle$ and
$\langle f_{0}   f_{0}  \rangle$. 
$\{f_{+}, f_0, f_{-}\}$ 
is an $SU(2)$ triplet, and transforms in the spin-1 
representation of $SU(2)$.

\subsection{Single-trace operators from the self-dual 
components}
We focus on the single-trace operators 
of length $\LL$ that are composed of self-dual components
only

\be
{\cO}(x) = \tr \ll f_{A_{1}}(x) \cdots f_{A_{\LL}}(x) \rr \,.
\label{ops}
\ee 

\noindent Following \cite{Ferretti:2004ba}, at 1-loop level, in the planar 
limit, these operators mix only among themselves, as in Equation 
({\bf \ref{f2pt}}), and their mixing matrix is given by 

\be
\Gamma = \frac{\lambda}{48\pi^{2}}\sum_{l=1}^{\LL}
\ll
7 + 3 \vec S_{l} \cdot \vec S_{l+1} 
- 3 (\vec S_{l} \cdot \vec S_{l+1})^{2}
\rr  \,,
\label{Hamiltonian}
\ee

\noindent where $\lambda=g^{2} N_{c}$, and 
$\vec S_{l}$ are $SU(2)$ spin-1 generators,

\begin{equation}
\ll S^j f \rr_A = -i \epsilon_{j A B} f_{B} \,.
\end{equation}

\noindent Note that $\{ f_{+}, f_{0}, f_{-}\}$ are eigenstates 
of $S^{3}$ with eigenvalues $\{+1, 0, -1\}$, respectively. Since 
$\Gamma$ commutes with $\vec S^{2}$ (where 
$\vec S = \sum_{l=1}^{\LL} \vec S_{l}$ is the total spin), and 
$S^{3}$, all three operators can be diagonalized simultaneously. 
An eigenstate of $\Gamma$ is an operator of definite conformal 
dimension $\Delta = 2 \LL + \gamma$, where $\gamma$ is the 
corresponding eigenvalue.

\subsection{Operator product expansion of single-trace
operators}

Following \cite{Escobedo:2010xs,Foda:2011rr}, we normalize 
the operators of definite conformal dimension according to

\be  
\langle {\cO}_{i}(x_{i})\,  \bar {\cO}_{j}(x_{j}) \rangle 
\sim  
\ll 
{\cN}_{i} {\cN}_{j}
\rr^{\frac{1}{2}} 
\frac{\delta_{ij}}{|x_{ij}|^{\Delta_{i}+\Delta_{j}}} 
\ee

\noindent for $x_{ij} \equiv x_{i}-x_{j} \rightarrow 0$, 
where ${\cN}_{i}$ will be specified below in 
Equation ({\bf \ref{Gaudin}}). The OPE of a pair of these 
operators ${\cO}_{1}(x)$ and ${\cO}_{3}(x)$ is given by

\be
\qquad {\cO}_1(x_1)\, {\cO}_3(x_3) \sim 
\sum_{{\cO}_2}
\ll
\frac{{\cN}_1 {\cN}_3}{{\cN}_2}
\rr^{\frac{1}{2}} 
\frac{C_{132}}{|x_{13}|^{\Delta_1 + \Delta_3 - \Delta_2}}
\ 
{\cO}_2 (x)  
+ \ldots \,, 
\quad 
x = \frac{1}{2} (x_1 + x_3)\,,
\label{OPE}
\ee

\noindent for $x_{13} \rightarrow 0$, where the ellipsis denotes 
subleading corrections involving conformal descendants of ${\cO}_{2}$ 
\cite{Braun:2003rp}. The structure constants $C_{132}$ have 
a perturbative expansion in $\lambda$,

\be
N_{c}\, C_{132} = c_{132}^{(0)} + \lambda c_{132}^{(1)}+ \ldots \,,
\ee

\noindent In this note, we focus on the leading (tree-level) contribution 
$c_{132}^{(0)}$.

\subsection{{\it \lq Tailoring\rq} the structure constants}
\label{sec:tailoring}
Following \cite{Escobedo:2010xs}, we construct $c_{132}^{(0)}$ in 
four steps. 

\subsection*{Step 1} We map the length-$L_i$ single-trace operator 
${\cO}_{i}$ to an eigenstate 
$| {\cO}_{i} \rangle$ of a length-$L_{i}$ periodic spin-1 chain Hamiltonian 
$\cH_1$. 

\subsection*{Step 2} We {\it \lq split\rq} the spin chains into left 
and right subchains of lengths 
\footnote{\ We restrict the discussion to the {\it \lq non-extremal\rq} 
case where all $\LL_{i,l}, \LL_{i,r}>0$, for which there is no mixing 
with double-trace operators \cite{Escobedo:2010xs}.}

\begin{equation}
\LL_{i,l} = \frac{1}{2}\left(\LL_{i}+\LL_{j} - \LL_{k}\right), 
\quad 
\LL_{i,r} = \frac{1}{2}\left(\LL_{i}+\LL_{k} - \LL_{j}\right),
\label{subchainlengths}
\end{equation}

\noindent respectively, with $(i,j,k)$ in cyclic order.
We perform a corresponding split of the states,

\be
| {\cO}_{i}  \rangle = \sum_{a} | {\cO}_{i_{a}}  \rangle_{l} \otimes | 
{\cO}_{i_{a}}  \rangle_{r}   \,,
\label{splitting}
\ee 

\noindent where, roughly speaking, the sum is over all possible ways 
of distributing the component fields into the left and right subchains. 
(A more precise definition of this splitting, as well as a more accurate 
version of Equation ({\bf \ref{splitting}}), will be given below after 
introducing the Bethe ansatz.) Note that 
$| {\cO}_{i_{a}} \rangle_{l}$ and 
$| {\cO}_{i_{a}} \rangle_{r}$
are states of subchains with lengths $\LL_{i,l}$ and $\LL_{i,r}$, 
respectively. 

\subsection*{Step 3} We {\it \lq flip\rq} or ${\cF}$-conjugate the right 
kets into right bras

\be
| {\cO}_{i}  \rangle = \sum_{a} | {\cO}_{i_{a}}  \rangle_{l} \otimes | 
{\cO}_{i_{a}}  \rangle_{r}  
\rightarrow  
\sum_{a} | {\cO}_{i_{a}}  \rangle_{l} \otimes\,
  {}_{r} \langle {\cO}_{i_{a}}|   \,.
\label{flip}
\ee 

\noindent Given a pair of elementary fields $A$ and $B$ that are 
associated with the kets $|\Psi_{i}\rangle_{r}$ and 
$|\Psi_{i+1} \rangle_{l}$, respectively, the flipped state 
${}_{r}\langle \Psi_{i}|$ is defined such that

\be
\langle A B \rangle \sim {}_{r}\langle \Psi_{i}| \Psi_{i+1}\rangle_{l} 
\label{prescription}
\,.
\ee

\noindent In view of the fact that the only non-zero 2-point functions 
are between $f_{+}$ and $f_{-}$, and between two $f_{0}$ fields, the 
prescription ({\bf \ref{prescription}}) implies that 

\be
|f_{\pm}\rangle_{r} \rightarrow  {}_{r}\langle f_{\mp}|, \qquad 
|f_{0  }\rangle_{r} \rightarrow  {}_{r}\langle f_{0  }|  \,.
\label{Fconjugation}
\ee

\noindent Our convention is that 
$\langle f_{\pm}| f_{\pm}\rangle = 1$,
$\langle f_{ 0 }| f_{ 0 }\rangle = 1$,
while all other 2-point functions are zero.

\subsection*{Step 4} We construct the structure constants by taking scalar 
products of bra and ket states 
\cite{Escobedo:2010xs,Foda:2011rr}, to obtain 

\begin{equation}
c_{132}^{(0)} = {\cN}_{132}\sum_{a,b,c} 
{}_{r}\langle {\cO}_{2_{b}} | {\cO}_{1_{a}}  \rangle_{l}\
{}_{r}\langle {\cO}_{1_{a}} | {\cO}_{3_{c}}  \rangle_{l}\ 
{}_{r}\langle {\cO}_{3_{c}} | {\cO}_{2_{b}}  \rangle_{l} \,,
\label{prelim}
\end{equation}

\noindent where

\begin{equation}
{\cN}_{132} = 
\ll
\frac{L_{1} L_{2} L_{3}}
{
\langle {\cO}_{1}| {\cO}_{1} \rangle
\langle {\cO}_{2}| {\cO}_{2} \rangle
\langle {\cO}_{3}| {\cO}_{3} \rangle}
\rr^{\frac{1}{2}
} \,.
\label{N123def}
\end{equation}

\begin{figure}
\begin{centering}
\includegraphics[height=7cm]{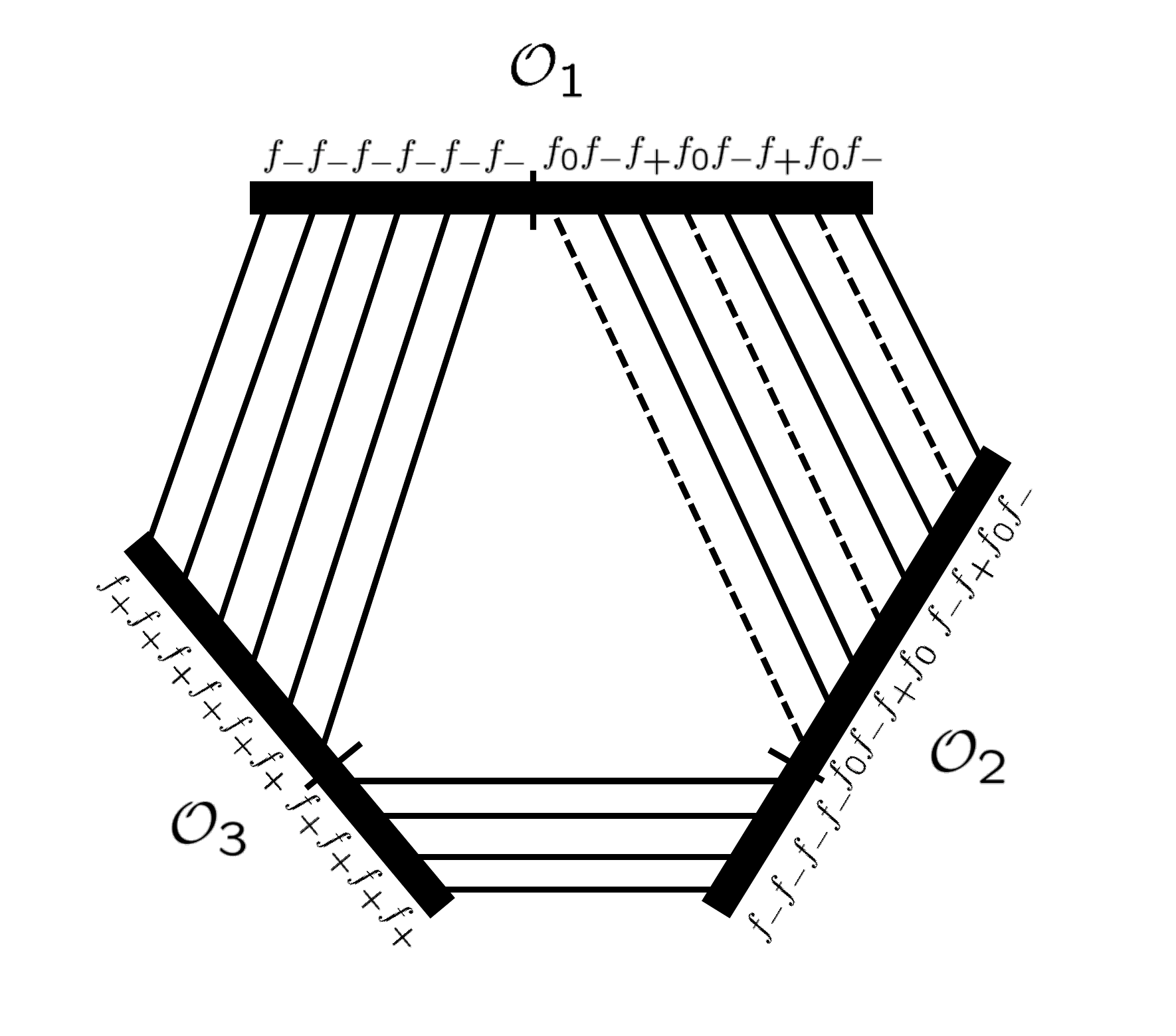}
\par\end{centering}
\caption{A configuration of 3-point functions with contractions
among the self-dual Yang-Mills fields 
(a solid line is $\langle f_{+}f_{-} \rangle$ and
a dotted line is $\langle f_{0}f_{0} \rangle$).
$\cO_3$ is chosen to consist of $f_{+}$ fields only, so it maps 
to a spin-chain reference state. This will be the case that can 
be evaluated in determinant form.}
\label{fig:fig1}
\end{figure}

\noindent This is represented graphically in 
Figure {\bf \ref{fig:fig1}}.  In order to further evaluate the 
expression ({\bf \ref{prelim}}) for the structure constants, it 
is necessary to have a more explicit construction of the states 
with definite conformal dimensions. To this end, we now turn to 
the Bethe ansatz.

\section{Algebraic Bethe ansatz}
\label{sec:BA}

\subsection{Diagonalizing the Hamiltonian $\cH_1$}
The 1-loop QCD mixing matrix $\Gamma$ ({\bf \ref{Hamiltonian}}) 
is identical 
to the Hamiltonian $\cH_1$ of an antiferromagnetic spin-1 chain, 
with periodic boundary conditions, that is integrable  
\cite{Zamolodchikov:1980ku}, 
and therefore can be diagonalized using the algebraic Bethe ansatz 
\cite{Kulish:1981gi, Kulish:1981bi, A.:1982zz, Babujian:1983ae}. 
The basic strategy to diagonalize $\cH_1$ is to diagonalize 
a transfer matrix 
$t^{(\frac{1}{2})}(u)$ that is constructed from a monodromy matrix 
with a 2-dimensional (that is, spin-$\frac{1}{2}$) auxiliary space. 
Although $t^{(\frac{1}{2})}(u)$ does not generate $\cH_1$ 
({\bf \ref{Hamiltonian}}), it is related by the fusion procedure 
to another transfer matrix $t^{(1)}(u)$ that is constructed from 
a monodromy matrix with a 3-dimensional (that is, spin-1) auxiliary 
space and that contains $\cH_1$ 
\cite{Kulish:1981bi,A.:1982zz, Babujian:1983ae}. 
By diagonalizing $t^{(\frac{1}{2})}(u)$, we diagonalize 
$t^{(1)}(u)$, $\cH_1$ and $\Gamma$, all in one go. 

\subsection{The $R$- and the monodromy matrices}
The transfer matrix $t^{(\frac{1}{2})}(u)$ can be constructed 
using the $6\! \times \! 6$ R-matrix

\begin{multline}
R^{(\frac{1}{2},1)}(u,v) = 
\\
\frac{1}{(u-v-\eta)}
\ll
\begin{array}{ccc|ccc}
    u-v+\eta  &              &                                  \\
              & u-v          &               &   \sqrt{2}\eta   \\
              &              &  u-v-\eta     & & \sqrt{2}\eta   \\
    \hline
              & \sqrt{2}\eta &               & u-v-\eta\\
              &              &  \sqrt{2}\eta &   &  u-v\\
              &              &               &   & & u-v+\eta 
\end{array} 
\rr \,,
\end{multline}

\noindent where eventually we shall set $\eta=i$. The matrix 
elements that are zero are left empty.  We regard 
$R^{(\frac{1}{2},1)}(u,v)$ as an operator acting on 
${\cC}^{2} \otimes {\cC}^{3}$. This R-matrix can be obtained 
by fusion \cite{Kulish:1981gi, Kulish:1981bi} 
from $R^{(\frac{1}{2},\frac{1}{2})}(u,v) = u-v + \eta {\cP}$, 
where ${\cP}$ is the permutation matrix on 
${\cC}^{2} \otimes {\cC}^{2}$, together with a {\it \lq gauge\rq} 
transformation that makes the matrix symmetric.

The (inhomogeneous) monodromy matrix is constructed from 
the $R$-matrix as
\footnote{In the sequel, we use different brackets to indicate 
the type of enclosed arguments. We write 
$f( x, y) $ when neither $x$ nor $y$ is a set of variables,  
$f\{x, y\}$ when both $x$ and $y$ are sets of variables,  
and 
$f[x, \{y\}]$ when $x$ is not a set of variables, but $y$ is.} 

\be
T^{(\frac{1}{2})}_{0}[u; \{z\}_{\LL}] = 
R_{01}^{(\frac{1}{2},1)}(u,z_{1}) \ldots 
R_{0\LL}^{(\frac{1}{2},1)}(u,z_{\LL})  \,,
\label{monodromy}
\ee

\noindent where we have introduced the inhomogeneities 
$\{z\}_{\LL}=\{z_{1}, \ldots, z_{\LL}\}$ for later convenience.  
The auxiliary space (labeled 0) is 2-dimensional, while each of 
the quantum spaces (labeled $1, \ldots, \LL$) are 3-dimensional
By tracing over the auxiliary space, we arrive at the (inhomogeneous) 
transfer matrix

\be
t^{(\frac{1}{2})}[u; \{z\}_{\LL}] = 
\tr_{0} T^{(\frac{1}{2})}_{0}[u; \{z\}_{\LL}] \,.
\ee 

\noindent It has the commutativity property 

\be
\left[
t^{(\frac{1}{2})}[u; \{z\}_{\LL}]  \,, 
t^{(\frac{1}{2})}[v; \{z\}_{\LL}]  
\right] = 0 \,,
\ee

\noindent by virtue of the fact that the R-matrix obeys the Yang-Baxter 
equation.

\subsection{Constructing the eigenstates}
The eigenstates of this transfer matrix can be readily obtained by 
algebraic Bethe ansatz: we define the operators $A, B, C, D$ by
\be
T^{(\frac{1}{2})}_{0}[u; \{z\}_{\LL}] =  
\ll 
\begin{array}{cc}
    A[u; \{z\}_{\LL}] & B[u; \{z\}_{\LL}] \\
    C[u; \{z\}_{\LL}] & D[u; \{z\}_{\LL}] 
    \end{array} 
\rr \,.
\label{ABCD}
\ee 

\noindent We also introduce the reference states with all spins up or 
all spins down,

\be
|0 \rangle_{\pm} = |f_{\pm} \rangle^{\otimes \LL} \equiv |f_{\pm}^{L} 
\rangle \,.
\label{reference}    
\ee 

\noindent These states are eigenstate of both $A[u; \{z\}_{\LL}]$ and 
$D[u; \{z\}_{\LL}]$,

\begin{multline}
A[u; \{z\}_{\LL}] |0 \rangle_{+} = 
\ll
\prod_{l=1}^{\LL}\frac{u-z_{l}+\eta}{u-z_{l}-\eta} 
\rr
|0 \rangle_{+} \,, 
\quad
D[u; \{z\}_{\LL}] |0 \rangle_{+} = |0 \rangle_{+} \,, 
\\
A[u; \{z\}_{\LL}] |0 \rangle_{-} =  |0 \rangle_{-} \,, 
\quad \quad \quad \quad \quad 
\quad \quad \quad 
D[u; \{z\}_{\LL}] |0 \rangle_{-} = 
\ll 
\prod_{l=1}^{\LL}\frac{u-z_{l}+\eta}{u-z_{l}-\eta} 
\rr
|0 \rangle_{-} \,.
\quad \quad 
\label{ADeigenvals}
\end{multline}

\noindent We note that

\be
B[u; \{z\}_{\LL}]^{\dagger} = - 
\ll
\prod_{l=1}^{\LL}\frac{u^{*} - z_{l}^{*}-\eta}{u^{*} - 
z_{l}^{*}+\eta}
\rr
C[u^{*}; \{z^{*}\}_{\LL}] \,, 
\label{Bdagger}
\ee 

\noindent where we have used $\eta = i$, and $*$ denotes complex 
conjugation. Choosing $|0 \rangle_{+}$ as the reference state, 
one finds that the states 

\be
| \{ u \}_{N} \rangle_{+} =  
\ll
\prod_{j=1}^{N} B[u_{j}; \{z\}_{\LL}] 
\rr
|0 \rangle_{+}
\label{Bethestates}
\ee

\noindent are eigenstates of the transfer matrix 
$t^{(\frac{1}{2})}[u; \{z\}_{\LL}]$ provided that 
$\{u\}_{N} =\{ u_{1}, \ldots, 
u_{N}\}$ are distinct and satisfy the spin-1 Bethe equations

\be
\prod_{l=1}^{\LL}\frac{u_{j}-z_{l}+\eta}{u_{j}-z_{l}-\eta} = 
\prod_{k=1 \atop k\ne j}^{N} 
\frac{u_{j}-u_{k}+\eta}{u_{j}-u_{k}-\eta} \,.
\label{BAE}
\ee 

\noindent In the homogeneous limit $z_{l}=0$, these states are 
eigenstates of $\cH_1$ ({\bf \ref{Hamiltonian}}) with eigenvalues 
(anomalous dimensions) \cite{Ferretti:2004ba}

\be
\gamma = \frac{\lambda}{48\pi^{2}}
\ll
7 \LL-\sum_{k=1}^{N} 
\frac{12}{u_{k}^{2}+1} 
\rr \,.
\ee 

\noindent The conformal dimensions are therefore given by 
$\Delta = 2\LL + \gamma$. The Bethe eigenstates 
({\bf \ref{Bethestates}}) 
are $SU(2)$ highest-weight states, with spin

\be
s=s^{3}=\LL-N \,,
\label{spin}
\ee 

\noindent and therefore $N \le \LL$. If we choose 
$|0 \rangle_{-}$ as the reference state, then the Bethe states 
are given by

\be
| \{ u \}_{N} \rangle_{-} =  
\ll
\prod_{j=1}^{N} C[u_{j}; \{z\}_{\LL}] 
\rr
|0 \rangle_{-}
\label{Bethestates-} \,,
\ee

\noindent which are lowest-weight states, with $s=-s^{3}=\LL-N$, 
so again $N \le \LL$.

In order to properly define the splitting of states 
({\bf \ref{splitting}}), we follow \cite{Escobedo:2010xs} 
and split the monodromy matrix ({\bf \ref{monodromy}}),

\be
T^{(\frac{1}{2})}_{0}[u; \{z\}_{\LL}] = 
T^{(\frac{1}{2})}_{0, l}[u; \{z\}_{\LL_{l}}]\, 
T^{(\frac{1}{2})}_{0, r}[u; \{z\}_{\LL_{r}}]\,, 
\ee

\noindent where

\be 
T^{(\frac{1}{2})}_{0, l}[u; \{z\}_{\LL_{l}}] &=& 
R_{01}^{(\frac{1}{2},1)}(u,z_{1}) \ldots 
R_{0\LL_{l}}^{(\frac{1}{2},1)}(u,z_{\LL_{l}})  \,, \non \\
T^{(\frac{1}{2})}_{0, r}[u; \{z\}_{\LL_{r}}] &=& 
R_{0,\LL_{l}+1}^{(\frac{1}{2},1)}(u,z_{\LL_{l}+1}) \ldots 
R_{0\LL}^{(\frac{1}{2},1)}(u,z_{\LL})  \,,
\ee 

\noindent and $\{z\}_{\LL_{l}} = \{z_{1}, \ldots, z_{\LL_{l}} \}$, 
$\{z\}_{\LL_{r}} = \{z_{\LL_{l}+1}, \ldots, z_{\LL} \}$.
Correspondingly, 

\begin{multline}
\ll 
\begin{array}{cc}
    A[u; \{z\}_{\LL}] & B[u; \{z\}_{\LL}] \\
    C[u; \{z\}_{\LL}] & D[u; \{z\}_{\LL}] 
    \end{array} 
\rr =
\\
\ll 
\begin{array}{cc}
    A_{l}[u; \{z\}_{\LL_{l}}] & B_{l}[u; \{z\}_{\LL_{l}}] \\
    C_{l}[u; \{z\}_{\LL_{l}}] & D_{l}[u; \{z\}_{\LL_{l}}] 
    \end{array} 
    \rr 
\ll 
\begin{array}{cc}
    A_{r}[u; \{z\}_{\LL_{r}}] & B_{r}[u; \{z\}_{\LL_{r}}] \\
    C_{r}[u; \{z\}_{\LL_{r}}] & D_{r}[u; \{z\}_{\LL_{r}}] 
    \end{array} 
    \rr  \,.
\end{multline} 

\noindent In particular,

\be
B[u; \{z\}_{\LL}] &=& 
A_{l}[u; \{z\}_{\LL_{l}}]\, 
B_{r}[u; \{z\}_{\LL_{r}}] + 
B_{l}[u; \{z\}_{\LL_{l}}]\,  
D_{r}[u; \{z\}_{\LL_{r}}]  \,, \non \\
C[u; \{z\}_{\LL}] &=& 
C_{l}[u; \{z\}_{\LL_{l}}]\, 
A_{r}[u; \{z\}_{\LL_{r}}] + 
D_{l}[u; \{z\}_{\LL_{l}}]\,  
C_{r}[u; \{z\}_{\LL_{r}}] \,.
\ee

\noindent The ${\cF}$-conjugation ({\bf \ref{Fconjugation4}}) 
implies that

\be
B_{r}[u; \{z\}_{\LL_{r}}] | f_{+}^{L_{r}} 
\rangle_{r} &\rightarrow& {}_{r}
\langle 
f_{-}^{L_{r}}| 
B_{r}[u; \{z\}_{\LL_{r}}] \,, \non \\
C_{r}[u; \{z\}_{\LL_{r}}] | f_{-}^{L_{r}} 
\rangle_{r} &\rightarrow&
{}_{r}\langle f_{+}^{L_{r}}| 
C_{r}[u; \{z\}_{\LL_{r}}] \,.
\ee

\section{Evaluating the structure constants}
\label{sec:structureconstants}

\subsection{3-point functions with three non-BPS-like 
operators in sum form} 
We start with the general case where all three composite 
operators $\cO_i$, $i \in \{1, 2, 3\}$ are non-BPS-like 
(they are not of highest or lowest conformal dimension), 
so they map to Bethe eigenstates $| \cO_i \rangle$ that 
are not spin-chain reference states, and that can be 
split into left and right parts as

\begin{multline}
|{\cO}_{i} \rangle = 
\\
\prod_{j=1}^{N_i}
\ll
A_l [u_{i, j}, \{z_{L_{i,l}}\}] 
\ 
B_r [u_{i, j}, \{z_{L_{i,r}}\}] +
D_r [u_{i, j}, \{z_{L_{i,r}}\}] 
\ 
B_l [u_{i, j}, \{z_{L_{i,l}}\}]
\rr
|f_{+}^{L_{i,l}}
\rangle_{l}
\otimes
|f_{+}^{L_{i,r}}\rangle_{r} 
=
\\
\sum_{\alpha_i \cup {\bar\alpha_i} =
\{ u_i \}_{N_i}} 
H_i \{ \alpha_i, \bar\alpha_i \}
| {\cO}_{i,    \alpha_i} \rangle_l 
\otimes 
| {\cO}_{i,\bar\alpha_i} \rangle_r 
\label{Oigen}
\end{multline}


\noindent where 

\begin{equation}
|{\cO}_{i, \alpha_i} \rangle_l =
\ll
\prod_{j \in     \alpha_i} B_l [u_{i, j}, \{z\}_{L_{i,l}}] 
\rr
| f_+^{L_{i, l}}
\rangle_{l}, 
\quad
|{\cO}_{i, \bar\alpha_i} \rangle_r =
\ll
\prod_{j \in \bar\alpha_i} B_r [u_{i, j}, \{z\}_{L_{i,r}}] 
\rr
| f_+^{L_{i, r}}
\rangle_{r}, 
\end{equation}

\noindent the coefficients 
$H_i \{ \alpha_i, \bar\alpha_i \}$ are computed from Equation 
({\bf \ref{ADeigenvals}}) to be

\begin{equation}
H_i \{ \alpha_i, \bar\alpha_i \} = 
\prod_{u_{i, j} \in \alpha_i}
\prod_{z_k      \in \{z\}_{L_{i,l}}} 
\frac{ u_{i, j} - z_k + \eta }{ u_{i, j} - z_k - \eta } \,,
\end{equation}

\noindent and $\{ u_i \}_{N_{i}}$ satisfy the Bethe 
equations ({\bf \ref{BAE}}) with $L=L_{i}$. 
Under ${\cF}$-conjugation, this operator becomes

\begin{equation}
|{\cO}_i \rangle
\to
\sum_{
\alpha_i \cup {\bar\alpha_i} = \{ u_i \}_{N_i}} 
H_i \{\alpha_i, \bar\alpha_i\}
| {\cO}_{i,\alpha_i} \rangle_l \otimes { }_r\langle 
{\cO}_{i,{\bar\alpha_i}}| \,,
\label{OiFconj}
\end{equation}

\noindent where

\begin{equation}
     { }_r\langle 
{\cO}_{i,{\bar\alpha_i}}| =
{}_r\langle f_{-}^{L_{i,r}}|
\ll
\prod_{j\in{\bar\alpha_i}} B_r[u_{i, j}, \{z\}_{L_{i,r}}] 
\rr \,.
\end{equation}

Substituting the above expressions into Equation 
({\bf \ref{prelim}}), we obtain the following sum expression for 
the structure constant of the 3-point function with three 
non-BPS-like operators

\begin{multline}
c_{132} = 
\\
\lim_{z_l \rightarrow 0}
{\cN}_{132} 
\sum_{\alpha_i \cup {\bar\alpha_i} = \{ u_i \}_{N_i}}
\ll
\prod_{i=1}^{3} H_i \{ \alpha_i, \bar\alpha_i \} 
\rr
{ }_r\langle {\cO}_{2, \bar\alpha_2} | {\cO}_{1, \alpha_1} \rangle_l
\ 
{ }_r\langle {\cO}_{1, \bar\alpha_1} | {\cO}_{3, \alpha_3} \rangle_l
\ 
{ }_r\langle {\cO}_{3, \bar\alpha_3} | {\cO}_{2, \alpha_2} \rangle_l 
\,,
\label{general-product}
\end{multline}

\noindent
where each of the three factors of type 
$
{ }_r\langle 
{\cO}_{i+1,{\bar\alpha_{i+1}}} | {\cO}_{i,\alpha_i}
\rangle_l
$
in the summand is a generic scalar product as in Equation
({\bf \ref{korepin-sum-spin-1/2}}),  
subject to the conditions in Equations ({\bf \ref{inhomospin1}}), 
({\bf \ref{aspin1}}). 

\subsection{The structure of the sum form in Equation 
({\bf \ref{general-product}})}
The non-BPS-like operator $\cO_i$, $i \in \{1, 2, 3\}$, 
is composed of the operators 
$\{f_{+},    f_{0},    f_{-}   \}$ with multiplicities 
$\{n_{i, +}, n_{i, 0}, n_{i, -}\}$,
such that $n_{i, +} + n_{i, 0} + n_{i, -}$$=$$ L_i$ 
and $2n_{i,-} + n_{i,0}=N_{i}$.
Splitting $\cO_i$ into 
a  left-part $\cO_{i, l}$ of length $L_{i, l}$, and 
a right-part $\cO_{i, r}$ of length $L_{i, r}$, 
$L_{i, l} + L_{i, r} = L_i$, the operators $\{f_+, f_0, f_-\}$
can be on either part, such that

\begin{equation}
n_{i, +}^l + n_{i, +}^r = n_{i, +}, \qquad 
n_{i, 0}^l + n_{i, 0}^r = n_{i, 0},  \qquad
n_{i, -}^l + n_{i, -}^r = n_{i, -}, 
\end{equation}

\noindent where $n_{i, +}^l$ is the number of $f_+$-operators 
on the left-part of $|\cO_i \rangle$, {\it etc.} 

Let us consider one type of these operators, for example $f_+$, 
to be a reference state operator, in the sense that if all 
elementary operators in a single-trace operator $\cO$ are 
of type $f_+$, then $\cO$ maps to a spin-chain reference state. 
In that case, the other two operators, $f_0$ and $f_-$, become 
{\it \lq excitations\rq}
\footnote{\ Either $f_+$ or $f_-$ can be chosen as a reference 
state operator, as we will see in the sequel.}. 
Since the total number of elementary operators in $\cO_i$, 
$i \in \{1, 2, 3\}$ is fixed, one can think of 
single-trace operators that are {\it not eigenstates} of the 
mixing matrix $\Gamma$, but whose weighted sum is a single-trace  
$\cO_i$ that is an eigenstate, as labeled by the positions 
of the excitations in the trace.

The crucial point is that, while the lengths of the left- and 
right-parts are fixed once and for all 
\footnote{\ This follows from the fact that the lengths 
$L_i$, $i \in \{1, 2, 3\}$ are fixed as initial conditions, 
and the lengths of the left and right parts are fixed from 
Equation ({\bf \ref{subchainlengths}}).}, 
the distribution of the excitations on the left and the right 
parts of $\cO_i$ is not fixed. This means that the sum in 
Equation ({\bf \ref{general-product}}) is over all possible 
distributions of excitations in $\cO_i$, $i \in \{1, 2, 3\}$ 
over its left and right parts, subject to the conditions

\begin{equation}
n_{i, +} = n_{i+1, -}, \qquad
n_{i, -} = n_{i+1, +}, \qquad
n_{i, 0} = n_{i+1, 0}, \qquad 
i + 3 \equiv i \,.
\end{equation}

In spin-chain terms, the action of the Bethe operators on 
a reference state, that consists of one type of operators, 
generates excitations of both types. Thus every state 
$\cO_i$, $i \in \{1, 2, 3\}$ that is not BPS-like will 
consist of all three types $\{f_+, f_0, f_-\}$, and we 
need to sum over all possible positions of 
$\{f_+, f_0, f_-\}$ in $\cO_i$. The result is that 
{\bf 1.} The sum over partitions in 
Equation ({\bf \ref{general-product}}) is computationally 
non-trivial, particularly when the number of Bethe roots 
involved is not small; and as mentioned above, 
{\bf 2.} Each of the three factors of type 
${ }_r\langle {\cO}_{i+1,{\bar\alpha_{i+1}}} | 
              {\cO}_{i,       \alpha_{i  } } \rangle_l $
in the summand is a generic scalar product as in Equation
({\bf \ref{korepin-sum-spin-1/2}}),  
subject to the conditions in Equation ({\bf \ref{inhomospin1}}). 
This is a complicated expression.

To reduce the complexity of the sum form in Equation ({\bf
\ref{general-product}}) and obtain a computationally tractable
expression, which in our case is a determinant, we choose one of
the operators to be BPS-like so that it maps to a spin-chain reference
state.  We will choose $\cO_3$ to be BPS-like.

\subsection{3-point functions with one BPS-like state in 
determinant form}

Choosing ${\cO}_{3}$ to consist of $f_{+}$-operators only,
the corresponding state is 

\be
| {\cO}_{3}  \rangle = |f_{+}^{\LL_{3}} \rangle =
|f_{+}^{\LL_{3,l}} \rangle_{l} \otimes  
|f_{+}^{\LL_{3,r}} \rangle_{r} \rightarrow 
|f_{+}^{\LL_{3,l}} \rangle_{l} \otimes {}_{r}\langle 
f_{-}^{\LL_{3,r}}| \,,
\label{O3}
\ee 

\noindent where $\LL_{3,l}$ and $\LL_{3,r}$ are given by 
(\ref{subchainlengths}). 
Evidently, since there is only one way to split this state, 
no summation is necessary. We write the algebraic Bethe state for 
operator ${\cO}_{1}$ as in ({\bf \ref{Oigen}}) with $i=1$, and we define the 
corresponding parameters $u_{j} \equiv u_{1,j}$,
%
%
which satisfy the Bethe equations 
({\bf \ref{BAE}}) with $L = L_1$. Under $\cF$-conjugation, this 
state becomes ({\bf \ref{OiFconj}}) with $i=1$.
%
%
%
%
\noindent However, we write the state corresponding to the operator 
${\cO}_2$ instead as 

\be
|{\cO}_{2}\rangle\to\sum_{\beta \cup {\bar\beta} = 
\{ v \}_{N_2}} H_2 \{\beta, \bar\beta \}
|{\cO}_{2,\beta}\rangle_l
\otimes 
{ }_r\langle {\cO}_{2,{\bar\beta}}|\,,
\ee

\noindent with

\be
\qquad \quad  |{\cO}_{2,\beta}\rangle_l =
\ll
\prod_{j\in\beta}C_l[v_j; \{z\}_{\LL_{2,l}}]
\rr
|f_{-}^{L_2, l} \rangle_l \,,
\qquad
{ }_r\langle {\cO}_{2,{\bar\beta}}|={}_r\langle f_{+}^{L_{2,r}}|
\ll
\prod_{j\in{\bar\beta}} C_r[v_j; \{z\}_{\LL_{2,r}}]
\rr \,,
\ee

\noindent where $\{ v \}_{N_{2}}$ satisfy the Bethe equations 
({\bf \ref{BAE}}) with $L=L_{2}$. Having chosen to construct the Bethe states 
for ${\cO}_1$ with the reference state $|0 \rangle_{+}$, it is necessary 
to construct the Bethe states for ${\cO}_2$ with the reference state 
$|0 \rangle_{-}$.
We now insert these results into Equation ({\bf \ref{prelim}}) to get

\begin{multline}
c_{132} =
\lim_{z_l \rightarrow 0} {\cN}_{132} 
 \sum_{
      \beta  \cup { \bar\beta} = \{ v \}_{N_2}
\atop \alpha \cup {\bar\alpha} = \{ u \}_{N_1}
}
H_1 \{ \alpha, \bar\alpha \}\, H_2 \{ \beta, \bar\beta \} \ 
{ }_r\langle {\cO}_{2, { \bar\beta}} | {\cO}_{1,\alpha} \rangle_l \
{ }_r\langle {\cO}_{1, {\bar\alpha}} |  f_{+}^{L_{3,l}} \rangle_l\
{ }_r\langle f_{-}^{L_{3, r}}        | {\cO}_{2, \beta} \rangle_l 
\\
=
\lim_{z_{l}\rightarrow 0}
{\cN}_{132}\, H_2 \{ \varnothing, \{ v \}_{N_2} \} 
\sum_{\alpha \cup {\bar\alpha} =  \{ u \}_{N_1} }
H_1 \{\alpha, \bar\alpha \} \
{ }_r\langle {\cO}_2 |{\cO}_{1, \alpha} \rangle_l \
{ }_r\langle {\cO}_{1, \bar\alpha} | f_{+}^{L_{3, l}} \rangle_l \
{ }_r\langle f_{-}^{L_{3, r}}| f_{-}^{L_{2,l}} \rangle_l
\label{c132more} \,.
\end{multline}

\noindent In passing to the second line, we have made use of the 
fact that the expression vanishes unless the set $\beta$ contains 
no Bethe roots, and we defined 

\be
{ }_r\langle{\cO}_{2}| \equiv \langle f_{+}^{L_{2,r}}|
\ll
\prod_{j=1}^{N_2} C_r [v_j; \{z\}_{\LL_{2,r}}] 
\rr \,.
\ee

\noindent With the help of Equation ({\bf \ref{ADeigenvals}}), 
we see that

\be
H_2 \{\varnothing, \{ v \}_{N_2} \} = 
\prod_{j=1}^{N_2}
\prod_{l=1}^{\LL_{2, l}} \frac{v_j -z_l +\eta} {v_j -z_l -\eta} 
\ee

\noindent becomes equal to 1 in the homogeneous limit, $z_l = 0$, 
by virtue of the zero-momentum constraint 

\be
\prod_{j=1}^{N_{2}}\frac{v_{j}+\eta}{v_{j}-\eta} = 1\,,
\ee 

\noindent which arises from the cyclicity of the trace in 
${\cO}_{2}$\ \footnote{
Note that this argument can be used only when all Bethe 
roots of an original unsplit eigenstate belong to the same 
part after splitting. This is the case for the eigenstate 
$| \cO_2\rangle$ in the 3-point function with one BPS-like
state. 
In particular, the same argument cannot be used to simplify 
the $H_i$ coefficients, $i \in \{1, 2, 3\}$, 
in Equation ({\bf \ref{general-product}}). This is because 
in the 3-point function with three non-BPS-like states, 
each state is split into a right part and a left part, 
and the Bethe roots can appear on either part. But neither 
part satisfies cyclicity on its own and the zero-momentum 
constraint cannot be used.}.
The remaining sum over partitions in Equation 
({\bf \ref{c132more}}) can be performed by using 
${ }_r\langle {\cO}_{1,{\bar\alpha}}|f_{+}^{L_{3,l}}\rangle_l=
{ }_l\langle f_{-}^{L_{3,l}}|{\cO}_{1,{\bar\alpha}}\rangle_r$.
Noting also that ${ }_r\langle 
f_{-}^{L_{3,r}}|f_{-}^{L_{2,l}}\rangle_l=1$, we obtain

\begin{multline}
c_{132}= \lim_{z_{l}\rightarrow 0}
{\cN}_{132}\, 
\sum_{\alpha \cup {\bar\alpha} =\{ u \}_{N_{1}}}
H_1 \{\alpha, \bar\alpha \}\
{}_r\langle {\cO}_{2}|{\cO}_{1,\alpha}\rangle_l\
{ }_l\langle f_{-}^{L_{3,l}}|{\cO}_{1,{\bar\alpha}}\rangle_r 
\\
= \lim_{z_{l}\rightarrow 0} {\cN}_{132}\,  
\ { }_l\langle f_{-}^{L_{3,l}}|\otimes\ {}_r\langle 
{\cO}_{2}|{\cO}_{1}\rangle.
\label{nopartition}
\end{multline}

We observe that this expression vanishes unless

\be
\LL_{2}-N_{2} = \LL_{1}+\LL_{3}-N_{1} \ge 0 \,.
\label{constraint}
\ee

\noindent Indeed, the factor 
${}_r\langle {\cO}_{2}|{\cO}_{1,\alpha}\rangle_l$ in 
the first line of Equation ({\bf \ref{nopartition}}) 
vanishes unless $|\alpha|$ (the number of Bethe roots in $\alpha$) is given 
by $|\alpha| = N_{2}$. It follows that $|\bar{\alpha}| = N_{1}-N_{2}$. 
Moreover, the two states in the factor 
${ }_l\langle f_{-}^{L_{3,l}}|{\cO}_{1,{\bar\alpha}}\rangle_r$ 
should have the same $S^{3}$ eigenvalue; hence,

\be
L_{1,r} - |\bar{\alpha}| = - L_{3,l} \,,
\ee

\noindent which then implies Equation ({\bf \ref{constraint}}). 
The sum over ${\cO}_{2}$ in Equation ({\bf \ref{OPE}}) can 
therefore be understood as the sum over all $\LL_{2}$ and 
$N_{2}$ satisfying the constraint ({\bf \ref{constraint}}).
The scalar product in the second line of 
Equation ({\bf \ref{nopartition}}) is 
a restricted Slavnov scalar product 

\be
c_{132}^{(0)} = {\cN}_{132}^{hom}\, 
 S^{hom}(\{u\}_{N_{1}}, 
\{v\}_{N_{2}})\,,
\label{finalresult}
\ee 

\noindent where $\{u\}_{N_{1}}, \{v\}_{N_{2}}$ are the Bethe 
roots corresponding to operators ${\cO}_{1}, {\cO}_{2}$, 
respectively. In Appendix {\bf \ref{sec:scalarproducts}} 
we obtain an expression  ({\bf \ref{restSlavnov}}) for 
the restricted Slavnov scalar product, which in 
the homogeneous limit $z_{l}\rightarrow 0$ becomes

\begin{multline}
\label{homrestSlavnov} 
S^{hom}(\{u\}_{N_{1}}, \{v\}_{N_{2}}) =
\\
\prod_{k=1}^{N_{2}} 
\ll 
\frac{v_{k}+\eta}{v_{k}-\eta} 
\rr^{\ll \frac{2L_1 - N_1 + N_2}{2}\rr}
\prod_{j>k}^{N_{1}}\frac{1}{u_{j}-u_{k}}
\prod_{j>k}^{N_{2}}\frac{1}{v_{j}-v_{k}}
\prod_{k=1}^{N_{2}}\frac{1}{
\ll
(v_{k}-\eta) v_{k}
\rr^{(N_{1}-N_{2})/2}} 
\\
\times 
\det
\ll
\begin{array}{cc}
    {\cM}_{ij} & 1 \le i \le N_{2}\,, \quad 1 \le j \le N_{1} \\
    \hline
    \Psi^{(i-1)}(u_{j},0) & 1 \le i \le (N_{1}-N_{2})/2 \,, 
    \quad 1 \le j \le N_{1} \\
    \hline
    \Psi^{(i-1)}(u_{j}+\eta,0) & 1 \le i \le (N_{1}-N_{2})/2 \,, 
    \quad 1 \le j \le N_{1}
    \end{array}
\rr \,,
\end{multline}

\noindent where

\begin{multline}
{\cM}_{ij} = \frac{\eta}{(u_{j}-v_{i})}
\ll
\prod_{m=1 \atop m\ne j}^{N_{1}}(v_{i}-u_{m}-\eta) -
\ll 
\frac{v_{i}-\eta}{v_{i}+\eta} 
\rr^{\LL_{1}}
\prod_{m=1 \atop m \ne j}^{N_{1}}(v_{i}-u_{m}+\eta) 
\rr \,,
\\
\Psi(u,z) = -\frac{1}{(u-z)(u-z-\eta)}
\prod_{j=1}^{N_{1}}(z-u_{j}) \,,
\quad
\Psi^{(j)}(u,z) =\frac{1}{j!}
\frac{\partial^{j}}{\partial z^{j}} \Psi(u,z) \,.
\end{multline}

Moreover, ${\cN}_{132}$ in Equation ({\bf \ref{N123def}}) 
is given by 

\be
{\cN}_{132} = 
\ll
\frac{\LL_{1}\LL_{2}\LL_{3}}{{\cN}_{1}{\cN}_{2}{\cN}_{3}}
\rr^{\frac{1}{2}} \,,
\label{N123}
\ee 

\noindent where ${\cN}_{i}$ are given by 
Equation ({\bf \ref{Gaudin}}). Indeed,

\begin{multline}
\langle {\cO}_{1}|{\cO}_{1}\rangle 
= 
{}_{+}\langle 0|\prod_{j=1}^{N_{1}} B[u_{j},\{z\}_{L_{1}}]^{\dagger} 
\prod_{j=1}^{N_{1}} B[u_{j},\{z\}_{L_{1}}] |0\rangle_{+} 
\\
= 
\ll
\prod_{j=1}^{N_{1}} 
\prod_{l=1}^{L_{1}}\frac{u_{j}^{*} - z_{l}^{*}-\eta}{u_{j}^{*} - 
z_{l}^{*}+\eta}
\rr
{}_{+}\langle 0|\prod_{j=1}^{N_{1}} C[u_{j}^{*},\{z^{*}\}_{L_{1}}]
\prod_{j=1}^{N_{1}} B[u_{j},\{z\}_{L_{1}}] |0\rangle_{+} \,,
\end{multline}

\noindent where we have used Equation ({\bf \ref{Bdagger}}). 
The prefactor becomes 1 in the homogeneous limit due to 
the zero-momentum constraint. Furthermore, the set of all 
Bethe roots $\{ u\}_{N_{1}}$ transforms into itself under 
complex conjugation. Hence, 

\be
\langle {\cO}_{1}|{\cO}_{1}\rangle^{hom} = 
\lim_{z_{l}\rightarrow 0} 
{}_{+}\langle 0|\prod_{j=1}^{N_{1}} C[u_{j},\{z\}_{L_{1}}]
\prod_{j=1}^{N_{1}} B[u_{j},\{z\}_{L_{1}}] |0\rangle_{+} = 
{\cN}_{1}^{hom} \,.
\ee 

\noindent Similar considerations apply to 
$\langle {\cO}_{2}|{\cO}_{2}\rangle$. 
Finally, we note that ${\cN}_{3}=1$.

\section{Discussion}\label{sec:discussion}

We have obtained a determinant expression for the tree-level OPE 
structure constants in planar QCD for operators of the type
({\bf \ref{ops}}), where one of them is BPS-like, 
see Equation ({\bf \ref{O3}}). 
Indeed, given $(\LL_{1}, N_{1})$ 
and $\LL_{3}$, the possible values of $(\LL_{2}, N_{2})$ are 
determined by Equation ({\bf \ref{constraint}}); then the 
corresponding Bethe equations ({\bf \ref{BAE}}) can be solved, 
and the structure constants $c_{132}^{(0)}$ can be efficiently 
computed using Equation ({\bf \ref{finalresult}}). 

In the QCD literature, operators of the form ({\bf \ref{ops}}) 
would be classified as {\it \lq chiral odd\rq}. While chiral-odd 
operators involving quark fields play an important role in certain 
hadronic scattering processes \cite{Jaffe:1991ra}, the purely gluonic 
chiral-odd operators that we have considered here (with no covariant 
derivatives) do not seem to have direct relevance to QCD phenomenology.

It would be interesting to generalize this work to operators with 
covariant derivatives, which are more relevant to phenomenology.
Such operators comprise the largest sector of QCD that is known to 
be integrable at one loop 
\cite{Beisert:2004fv, Braun:1998id, Braun:1999te, Belitsky:1999qh, Ahn:2007ky}.
Another challenge is to go to higher loops (see {\it e.g.} 
\cite{Belitsky:2005bu, Gromov:2012vu}).

\section*{Acknowledgments}
We thank G Korchemsky for valuable comments and help with
references, and M Wheeler for discussions.
CA and RN are grateful for the warm hospitality extended to them 
at the Nordita workshop 
{\it \lq Exact Results in Gauge-String Dualities\rq}, 
where some of this work was performed. CA also thanks University 
of Melbourne and University of Miami for hospitality.
This work was supported in part by the World Class University grant
R32-2009-000-10130-0 (CA), The Australian Research Council (OF), 
and by the National Science Foundation 
under Grant PHY-0854366 and a Cooper fellowship (RN).

\appendix

\section{Coordinate Bethe ansatz and ${\cF}$-conjugation}\label{sec:CBA}

In order to properly formulate ${\cF}$-conjugation in the algebraic Bethe 
ansatz formalism, it is necessary to first formulate it in the coordinate 
Bethe ansatz formalism. 

We begin by reviewing the coordinate Bethe ansatz for spin-1, which has 
been discussed in 
\cite{LimaSantos:1998te, Crampe:2010xx}.
For simplicity, we consider the homogeneous case $z_{l}=0$, and restrict 
to states with just two excitations, which are given by

\be
|\{u_{1}, u_{2}\}\rangle^{co} = \sum_{1\le n_{1} \le n_{2} \le L} 
\ll
e^{i(p_{1} n_{1} + p_{2} n_{2})} + S(p_{2}, p_{1})\,  
e^{i(p_{2} n_{1} + p_{1} n_{2})} 
\rr 
|n_{1}, n_{2} \rangle \,.
\label{COstate}
\ee

\noindent Here $|n_{1}, n_{2} \rangle$ is given by  
\cite{Crampe:2010xx}

\be
|n_{1}, n_{2} \rangle = e^{-}_{n_{1}} e^{-}_{n_{2}} |f_{+}^{L} 
\rangle \,, \qquad 
e^{-} = 
\ll 
\begin{array}{ccc}
    0 & 0 & 0\\
    2^{1/2} & 0 & 0   \\
    0 &  2^{-1/2} & 0
    \end{array} 
\rr \,,
\ee

\noindent and

\be
S(p_{2},p_{1})=\frac{u_{2}-u_{1}+i}{u_{2}-u_{1}-i}\,, \qquad e^{i 
p_{j}} = \frac{u_{j}+i}{u_{j}-i} \,.
\label{Smat}
\ee 

\noindent The expression ({\bf \ref{COstate}}) is almost the same 
as for the spin-$\frac{1}{2}$ case \cite{Escobedo:2010xs}, the main 
difference is that now the summation includes $n_{1} = n_{2}$.

We define ${\cF}$-conjugation by

\be
{\cF} \circ |n_{1}, n_{2} \rangle = 
\langle L+1-n_{2}, L+1-n_{1}| \  
\hat {\cC}^{\otimes L} \,,
\label{Fconjugation2}
\ee 

\noindent where

\be
\hat {\cC} = \ll \begin{array}{ccc}
     0 & 0 & 1 \\
     0 & 1 & 0 \\
     1 & 0 & 0 
    \end{array} \rr = \hat {\cC}^{\dagger} \,,
\ee 

\noindent which has the properties

\begin{equation}
\hat {\cC} |f_{\pm} \rangle = |f_{\mp} \rangle, 
\qquad  
\hat {\cC} |f_{0} \rangle = |f_{0} \rangle, 
\qquad  
\hat {\cC}^{2} = 1,  
\qquad
\hat {\cC}^{\otimes L} B(u)\, \hat {\cC}^{\otimes L} =  C(u) \,.
\end{equation}

\noindent The definition ({\bf \ref{Fconjugation2}}) is consistent 
with Equation ({\bf \ref{Fconjugation}}), and is a generalization 
of the definition for the spin-$\frac{1}{2}$ case 
\cite{Escobedo:2010xs}. It follows, as in the spin-$\frac{1}{2}$ 
case, that ${\cF}$-conjugation of the coordinate Bethe ansatz state 
({\bf \ref{COstate}}) is given by

\be
{\cF} \circ |\{u_{1}, u_{2}\}\rangle^{co} = 
e^{i(L+1)(p_{1}+p_{2})} S(p_{2},p_{1})\, {}^{co}\langle 
\{u_{1}^{*},u_{2}^{*}\}| \hat {\cC}^{\otimes L} \,,
\label{Fconjugation3}
\ee

\noindent where 
${}^{co}\langle 
\{u_{1},u_{2}\}| \equiv 
\ll 
|\{u_{1}, u_{2}\}\rangle^{co} 
\rr^{\dagger}$.

We now proceed to translate this result to the algebraic Bethe ansatz.
One can show that the algebraic and coordinate Bethe ansatz states are 
related (in our normalization) by

\be
|\{u_{1}, u_{2}\}\rangle^{al} = 
-\frac{(u_{1}-u_{2}+i)}{(u_{1}+i)(u_{2}+i)(u_{1}-u_{2})} |\{u_{1}, 
u_{2}\}\rangle^{co} \,,
\label{ABACBA}
\ee 

\noindent generalizing the known spin-$\frac{1}{2}$ result 
\cite{Ovchinnikov:2010vb, Escobedo:2010xs}.
The corresponding hermitian-conjugate result is 

\be
{}^{al}\langle 
\{u_{1},u_{2}\}| \equiv 
\ll 
|\{u_{1}, u_{2}\}\rangle^{al} 
\rr^{\dagger} = 
-\frac{(u_{1}^{*}-u_{2}^{*}-i)}
{(u_{1}^{*}-i)(u_{2}^{*}-i)(u_{1}^{*}-u_{2}^{*})}\, {}^{co}
\langle 
\{u_{1},u_{2}\}| \,,
\ee 

\noindent and therefore

\be
{}^{co}\langle \{u_{1}^{*},u_{2}^{*}\}| = - 
\frac{(u_{1}-i)(u_{2}-i)(u_{1}-u_{2})}{(u_{1}-u_{2}-i)}\, {}^{al}\langle 
\{u_{1}^{*},u_{2}^{*}\}| \,.
\label{ABACBAdag}
\ee

\noindent Using Equations 
({\bf \ref{ABACBA}}, {\bf \ref{Fconjugation3}}, {\bf \ref{ABACBAdag}}) 
and ({\bf \ref{Smat}}), we obtain

\be
{\cF} \circ |\{u_{1}, u_{2}\}\rangle^{al} = 
\prod_{j=1}^{2} 
\ll
\frac{u_{j}+i}{u_{j}-i}
\rr^{L}\, {}^{al}
\langle 
\{u_{1}^{*},u_{2}^{*}\}|  \hat {\cC}^{\otimes L} \,.
\ee 

\noindent Since 
$|\{u_{1}, u_{2}\}\rangle^{al} = B(u_{1}) B(u_{2}) 
|f_{+}^{L}\rangle $, with the help of 
Equation ({\bf \ref{Bdagger}}) we see that

\be
{}^{al}\langle 
\{u_{1}^{*},u_{2}^{*}\}| = 
\prod_{j=1}^{2}
\ll 
\frac{u_{j}-i}{u_{j}+i}
\rr^{L}\,  
\langle f_{+}^{L}| C(u_{1}) C(u_{2}) \,.
\ee

\noindent We conclude that ${\cF}$-conjugation of an algebraic Bethe 
ansatz state is given by

\be
{\cF} \circ 
\ll
B(u_{1}) B(u_{2}) |f_{+}^{L}\rangle 
\rr 
= 
\langle f_{+}^{L}| C(u_{1}) C(u_{2}) \hat 
{\cC}^{\otimes L}  = 
\langle f_{-}^{L}| B(u_{1}) B(u_{2}) \,.
\label{Fconjugation4}
\ee

\section{Scalar products}\label{sec:scalarproducts}

\subsection{Izergin's determinant}
To define the scalar product of two spin-chain states that 
are not eigenstates of the Hamiltonian, we need Izergin's
determinant expression 
\cite{Izergin:1981aa} for Korepin's 
{\it \lq domain wall partition function\rq}
\cite{Korepin:1982gg}.
For two sets of variables
$\{x\}$ and 
$\{y\}$ of cardinality 
$|x|$ $=$ 
$|y|$ $=$ $\ell$, Izergin's determinant expression 
$Z\{x, y\}$ is

\begin{equation}
Z\{x, y\}
=
\frac{
\prod_{i,j=1}^{\ell}
(x_i - y_j + \eta)
}
{
\prod_{1 \leq i < j \leq \ell}
(x_j - x_i)
(y_i - y_j)
}
\det
\ll
\frac{1}{(x_i - y_j + \eta)(x_i - y_j)} 
\rr_{1 \leq i, j \leq \ell} \,,
\label{izergin}
\end{equation}

\noindent where $\eta=\frac{i}{2}$.

\subsection{The generic scalar product in an XXX spin-$\frac{1}{2}$ 
chain}
For a length-$L$ periodic XXX spin-$\frac{1}{2}$ chain, we consider 
{\bf 1.} Two non-BPS-like states in the space of states of the spin 
chain,
$| \cO_i \{u\}\rangle$ and 
$| \cO_j \{v\}\rangle$,
$| u | = 
 | v | = N \leq L/2$,
that are {\it not} Bethe eigenstates of the Hamiltonian 
$\cH_{\frac{1}{2}}$, that is, 
$\{u\}$ and $\{v\}$ do {\it not} satisfy Bethe equations, and 
{\bf 2.} The set of all possible partitions of each of $\{u\}$ 
and $\{v\}$ into two disjoint subsets

\begin{align}
\{u\} = \{u_1\} \cup \{u_2\}, \quad
\{v\} = \{v_1\} \cup \{v_2\}, \quad
0 \leq |u_1| = |v_1| \leq N,        \quad 
0 \leq |u_2| = |v_2| \leq N \,,
\end{align}

\noindent where 
$\{u_1\}$ $ = $ $\{u_{1, 1}, u_{1, 2}, \cdots, u_{1, |u_1|} \}$, 
{\it etc.} 
Following \cite{Korepin:1982gg, Korepin:1993aa}, the scalar product 
$\langle 
\cO_j \{v\} | 
\cO_i \{u\} 
\rangle$, 
is 

\begin{multline}
\langle \cO_j \{v\} | \cO_i \{u\} \rangle
=
\sum_{
\{u_1\} \cup  \{u_2\}, 
\{v_1\} \cup  \{v_2\}
}
\ll
\prod_{ \{u_1\} }
a^{\frac{1}{2}} [u_1, \{z\}_L] 
\prod_{ \{v_2\} }
a^{\frac{1}{2}} [v_2, \{z\}_L]
\rr
\\
\times
\ll
\prod_{i=1}^{|v_1|}
\prod_{j=1}^{|v_2|}
f ( v_{1, i}, v_{2, j} ) 
\rr
\ll
\prod_{i=1}^{|u_2|}
\prod_{j=1}^{|u_1|}
f ( u_{2, i}, u_{1, j} ) 
\rr
Z\{ v_1, u_1 \}
Z\{ u_2, v_2 \} \,,
\label{korepin-sum-spin-1/2}
\end{multline}

\noindent where the sum is over all partitions of 
$\{u\}$ and $\{v\}$ into two disjoint subsets, and

\begin{equation}
a^{\frac{1}{2}} [x, \{z\}_L] = \prod_{i=1}^{L} 
\frac{x - z_i + \eta}
     {x - z_i       }, 
\quad
f(x_i, y_j) = \frac{x_i - y_j + \eta}{x_i - y_j} \,.
\end{equation}

\subsection{The generic scalar product in an XXX spin-1 chain}
In the case of a length-$L$ XXX spin-1 chain case, the generic 
scalar product has the same form as in 
Equation ({\bf \ref{korepin-sum-spin-1/2}}), 
but with the following extra conditions. 
{\bf 1.} We start from a spin-$\frac{1}{2}$ chain with $2L$ sites. 
{\bf 2.} We set the inhomogeneities 

\begin{equation}
z_{2i + 2} = z_{2i + 1} + \eta, \quad i \in \{0, 1, \cdots, (L-1) \} 
\,,
\label{inhomospin1}
\end{equation}

\noindent as required by fusion. 
{\bf 3.} We take the inhomogeneities $w_i$, $i \in \{1, 2, \cdots, L\}$ 
of the $L$-sites of the 
spin-1 chain to be those of the odd-indexed sites of the original 
spin-$\frac{1}{2}$ chain, $w_i = z_{2i - 1}$.
{\bf 4.} We change $a^{\frac{1}{2}} [x, \{z\}_L]$ to 
                   $a^1             [x, \{w\}_L]$ defined by

\begin{equation}
a^1 [x, \{w\}_L ] = \prod_{i=1}^{L} 
\frac{x - w_i + \eta}
     {x - w_i - \eta}, 
\quad 
\eta = \frac{i}{2} \,.
\label{aspin1}
\end{equation}

\noindent while all other factors remain unchanged as they have 
no dependence on the inhomogeneities. The result is the generic 
scalar product for the spin-1 chain.

\subsection{The Slavnov scalar product}

\begin{figure}
\begin{centering}
\includegraphics[height=5cm]{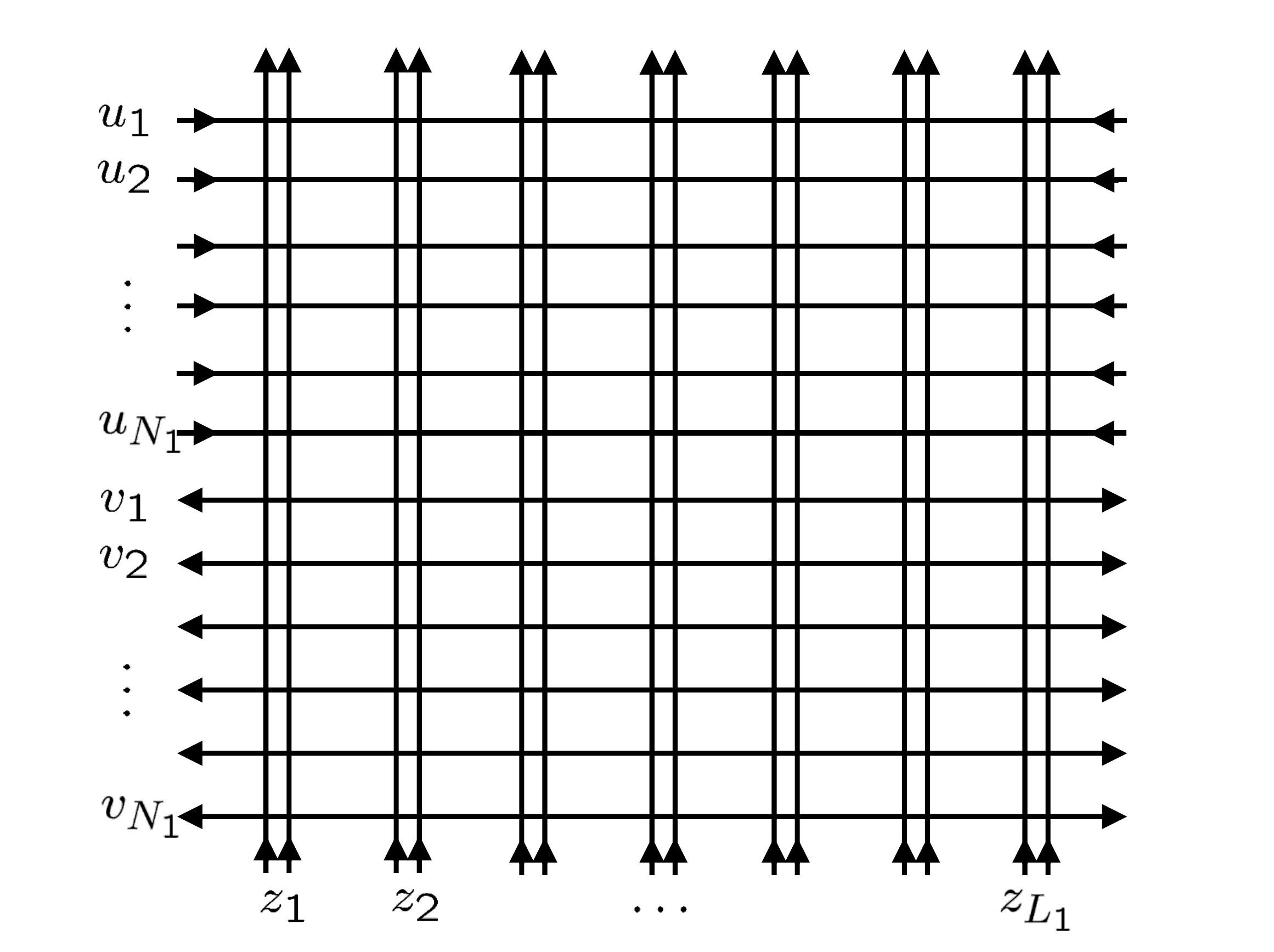}
\par\end{centering}
\caption{2D lattice configuration for the Slavnov determinant. 
Double vertical lines denote spin-1 quantum spaces with double 
up-arrows for the $f_{+}$ state. Horizontal lines with incoming 
spin-$\frac{1}{2}$ arrows denote $B$ operators, while those with 
outgoing arrows denote $C$ operators. If we impose 
$u_i=v_i\ (i=1,\ldots,N_1)$, then this configuration depicts the 
Gaudin norm.}
\label{fig:fig2}
\end{figure}

Let us first consider the matrix element 

\be
S_{N}(\{u\}_{N}, \{v\}_{N}, \{z\}_{\LL}) = 
\langle 0| \prod_{j=1}^{N} 
C[v_{j}; \{z\}_{\LL}]  \prod_{k=1}^{N} 
B[u_{k}; \{z\}_{\LL}] |0 \rangle 
\,,
\label{Slavnov}
\ee

\noindent where $\{u\}_{N} = \{ u_{1}, \ldots, u_{N}\}$ (but not 
necessarily $\{v\}_{N}= \{ v_{1}, \ldots, v_{N}\}$) satisfy the 
Bethe equations ({\bf \ref{BAE}}), and 
$|0 \rangle \equiv |0 \rangle _{+}$. 
In the 2-dimensional vertex-model description, this scalar product 
is represented by Figure {\bf \ref{fig:fig2}}. It follows from Slavnov 
\cite{Slavnov:1989aa} that this matrix element is given by
\footnote{\ We identify $-ic$ in \cite{Slavnov:1989aa} with $\eta$.}

\be
S_{N}(\{u\}_{N}, \{v\}_{N}, \{z\}_{\LL}) = 
\ll
\prod_{j>i}^{N} 
\frac{1}{v_{j}-v_{i}}\frac{1}{u_{i}-u_{j}} 
\rr
\det M_{lk} \,,
\label{Slavnov2}
\ee

\noindent where the $N \times N$ matrix $M_{lk}$ is given by

\be
M_{lk} = \frac{\eta }{u_{k}-v_{l}}
\ll
\prod_{m=1 \atop m\ne k}^{N}(v_{l}-u_{m}-\eta)
\prod_{j=1}^{\LL}\frac{v_{l}-z_{j}+\eta}{v_{l}-z_{j}-\eta}
-\prod_{m=1 \atop m\ne k}^{N}(v_{l}-u_{m}+\eta) 
\rr \,.
\label{Slavnov3}
\ee

\subsection{Gaudin norm}

For the special case that $\{ v_{i} \}$ coincide with $\{u_{i} \}$, 
the scalar product ({\bf \ref{Slavnov}}) reduces to the Gaudin norm 
\cite{Gaudin:1976aa, Gaudin:1981aa, Korepin:1982gg}

\begin{multline}
{\cN}(\{u\}_{N}, \{z\}_{\LL}) = 
\langle 0| 
\prod_{j=1}^{N} C[u_{j}; \{z\}_{\LL}]  
\prod_{k=1}^{N} B[u_{k}; \{z\}_{\LL}] 
|0  \rangle   
= 
\eta^{N} 
\ll 
\prod_{j \ne k}\frac{u_{j}-u_{k}-\eta}{u_{j}-u_{k}} 
\rr
\det \Phi' \,, 
\label{Gaudin}  
\end{multline}

\noindent where $\Phi'$ is an $N \times N$ matrix given by

\be 
\Phi'_{jk} = \frac{\partial}{\partial u_{k}} 
\log 
\ll
\prod_{l=1}^{\LL}\frac{u_{j}-z_{l}+\eta}{u_{j}-z_{l}-\eta} 
\prod_{m\ne j} \frac{u_{j}-u_{m}-\eta}{u_{j}-u_{m}+\eta} 
\rr  \,.
\ee

\subsection{Restricted Slavnov scalar product}

We now show how to restrict the Slavnov scalar product
({\bf \ref{Slavnov}})-({\bf \ref{Slavnov3}}) 
(with $N=N_{1}$ and $\LL=\LL_{1}$) 
to obtain Equation ({\bf \ref{restSlavnov}}).  
The basic trick 
\cite{Foda:2011rr, Kitanine:1999aa, Wheeler:2011xw} 
is to set the {\it \lq extra\rq} $v$-variables equal 
to inhomogeneities:

\begin{equation}
v_{N_1 -2j +1} = z_j, 
\qquad
v_{N_{1}-2j+2} = z_{j}+\eta, 
\qquad
j= 1, \ldots, \frac{1}{2}(N_{1}-N_{2}), \quad 
N_2 < N_1 \,.
\label{vvalues}
\end{equation}

\noindent However, since the expression ({\bf \ref{Slavnov3}}) 
for $M_{lk}$ then becomes singular, it is convenient to first 
change normalization. Using a tilde to denote quantities in 
the new normalization, we see that

\be
\tilde R^{(\frac{1}{2},1)}(u,v)
=
\alpha(u,v)\, R^{(\frac{1}{2},1)}(u,v)
\ee

\noindent implies that

\be
\tilde B[u; \{z\}_{\LL}] = 
\prod_{l=1}^{\LL}\alpha(u,z_{l})\, B[u; \{z\}_{\LL}] \,, 
\quad
\tilde C[u; \{z\}_{\LL}] = 
\prod_{l=1}^{\LL}\alpha(u,z_{l})\, C[u; \{z\}_{\LL}] \,.
\label{renBC}
\ee

\noindent Hence,

\begin{equation}
\tilde S_{N_{1}} \equiv 
\langle 0| \prod_{j=1}^{N_{1}} 
\tilde 
C[v_{j}; \{z\}_{\LL_{1}}]  
\prod_{k=1}^{N_{1}}\tilde B[u_{k}; \{z\}_{\LL_{1}}] 
|0 \rangle  
= \ll
\prod_{j=1}^{N_{1}}  
\prod_{l=1}^{\LL_{1}} 
\alpha(u_{j},z_{l})\alpha(v_{j},z_{l})\, 
\rr S_{N_{1}} \,.
\label{renSlavnov}
\end{equation}

\noindent We choose the normalization factor

\be
\alpha(u,v) = \frac{u-v-\eta}{u-v+\eta} \,,
\ee

\noindent which will avoid the singularity. Then 

\be
\tilde S_{N_{1}} = 
\ll
\prod_{j=1}^{  N_{1}}  
\prod_{l=1}^{\LL_{1}} 
\frac{u_{j}-z_{l}-\eta}{u_{j}-z_{l}+\eta}
\rr
\ll
\prod_{j>i}^{N_1} 
\frac{1}{v_{j}-v_{i}}\frac{1}{u_{i}-u_{j}}
\rr 
\det \tilde M_{lk} \,,
\label{renSlavnov2}
\ee

\noindent where

\be
\tilde M_{lk} = \frac{\eta }{(u_{k}-v_{l})}
\ll
\prod_{m=1 \atop m\ne 
k}^{N_1}(v_{l}-u_{m}-\eta)
-\prod_{j=1}^{\LL_1}\frac{v_{l}-z_{j}-\eta}{v_{l}-z_{j}+\eta}
\prod_{m=1 \atop m\ne k}^{N_1}(v_{l}-u_{m}+\eta) 
\rr \,.
\label{renSlavnov3}
\ee

\begin{figure}
\begin{centering}
\includegraphics[height=5.5cm]{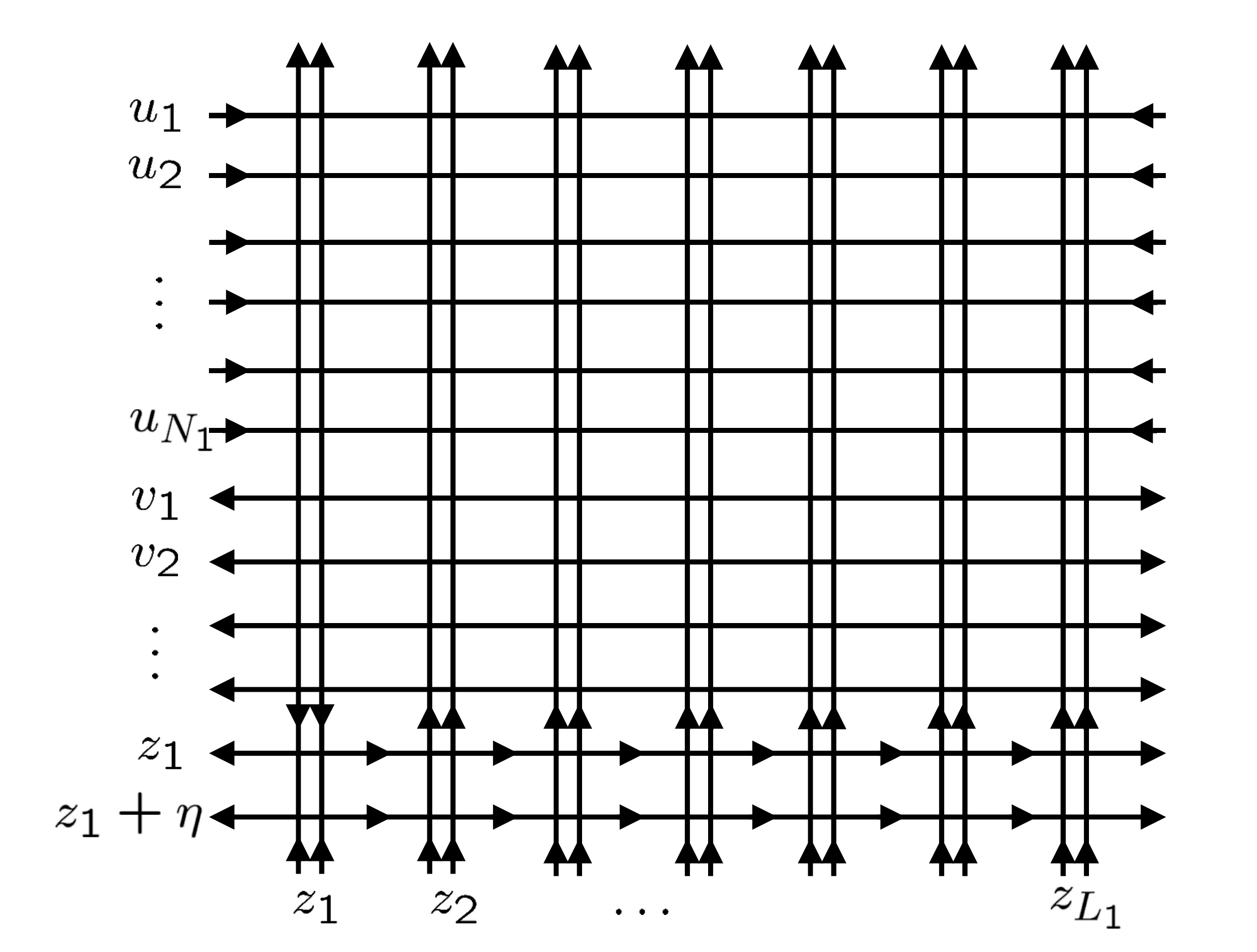}
\includegraphics[height=5.5cm]{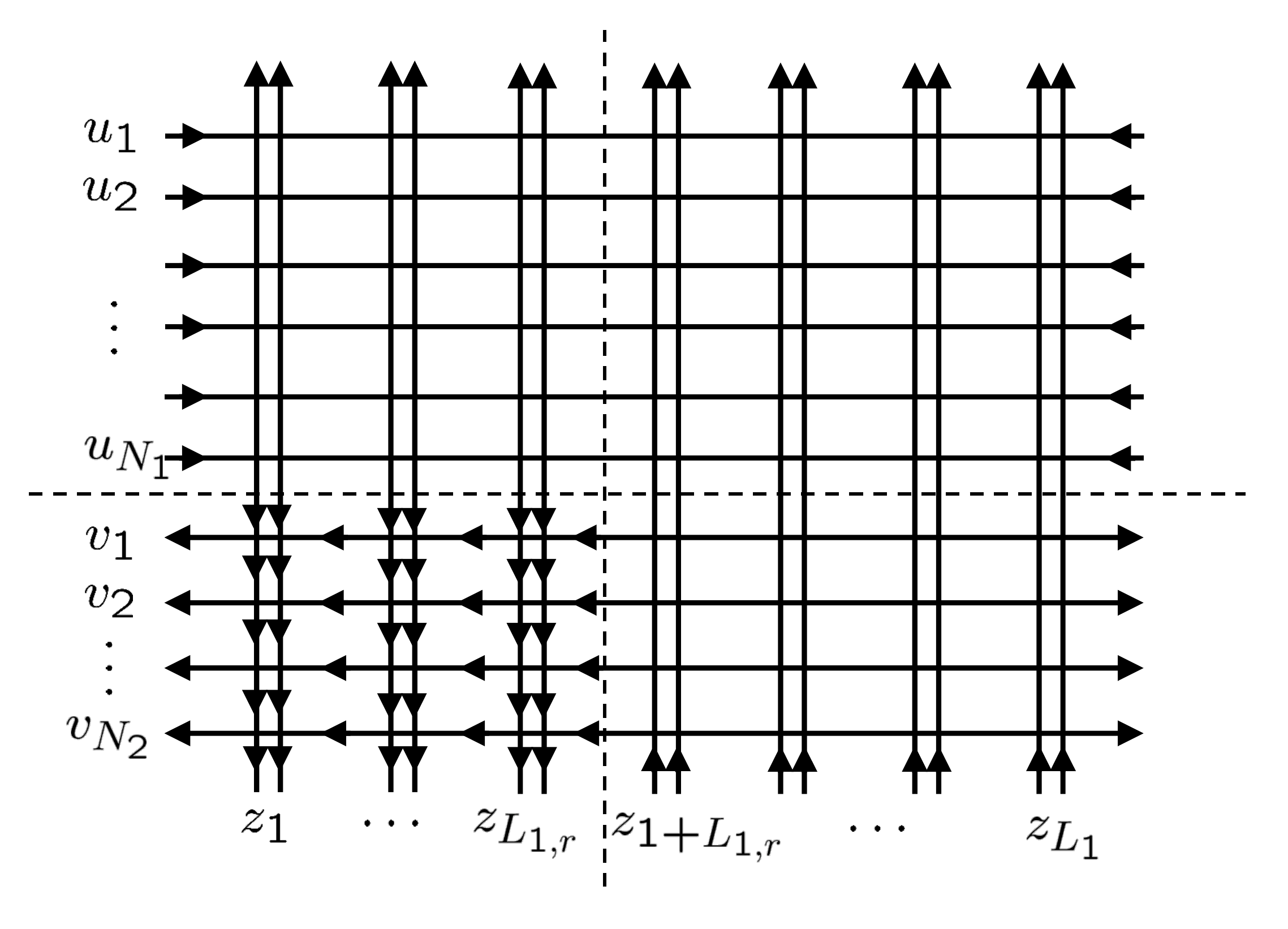}
\par\end{centering}
\caption{On the left, we {\it \lq freeze\rq} the two bottom rows 
of Figure {\bf \ref{fig:fig2}} by imposing 
Equation ({\bf \ref{vvalues}}). 
These two frozen rows are then eliminated. By repeating this 
procedure, we freeze out and eliminate the $N_1-N_2$ bottom 
rows, thereby obtaining the figure on the right. The spins in 
the bottom-left part of the remaining lattice are in fact 
completely fixed. After also removing this part, we obtain 
the restricted Slavnov determinant in 
Figure {\bf \ref{fig:fig4}}.
}
\label{fig:fig3}
\end{figure}

We are now ready to {\it \lq freeze\rq}, or {\it \lq restrict\rq}, 
the scalar product $\tilde S_{N_{1}}$ by setting 
$\{ v_{N_{2}+1}, $$\ldots, $$v_{N_{1}}\}$ to the values in Equation 
({\bf \ref{vvalues}}), to obtain $\tilde S_{restricted}$, which is 
(see Figure {\bf \ref{fig:fig3}})

\begin{multline}
\tilde S_{restricted} = 
\prod_{j=1}^{N_{1}}
\prod_{l=1}^{\LL_{1}}\frac{u_{j}-z_{l}-\eta}{u_{j}-z_{l}+\eta}
\prod_{N_{1}\ge j>k\ge 1}\frac{1}{u_{j}-u_{k}} 
\prod_{N_{2}\ge j>k\ge 1} \frac{1}{v_{j}-v_{k}} 
\\
\times 
\prod_{\frac{1}{2}(N_{1}-N_{2})\ge j>k\ge1} 
\frac{1}{(z_{j}-z_{k})^{2}(z_{j}-z_{k}-\eta)(z_{j}-z_{k}+\eta)} 
\\
\times 
\prod_{j=1}^{\frac{1}{2}(N_{1}-N_{2})}
\prod_{k=1}^{N_{2}}\frac{1}{(z_{j}-v_{k}+\eta)(z_{j}-v_{k})} 
\det {\cM}_{lk} \,,
\end{multline}

\noindent where ${\cM}_{lk}$ is an $N_{1} \times N_{1}$ matrix, 
which for $l \le N_{2}$ is the same as ({\bf \ref{renSlavnov3}}), namely

\begin{multline}
{\cM}_{lk} =
\frac{\eta }{(u_{k}-v_{l})}
\ll
 \prod_{m=1 \atop m\ne k}^{N_1}(v_{l}-u_{m}-\eta)
-\prod_{m=1 \atop m\ne k}^{N_1}(v_{l}-u_{m}+\eta) 
\prod_{j=1}^{\LL_1}
\frac{v_{l}-z_{j}-\eta}{v_{l}-z_{j}+\eta}
\rr
\,, \ l \le N_2 \,;
\label{restSlavnov2a} 
\end{multline}

\noindent and for $l > N_{2}$,

\be
{\cM}_{N_{2}+2j-1,k} &=&  \frac{1 }{(u_{k}-z_{j})}
\ll
\prod_{n=1 \atop n\ne k}^{N_1}(z_{j}-u_{n}-\eta) 
-\prod_{l=1}^{\LL_{1}}\frac{z_{j}-z_{l}-\eta}{z_{j}-z_{l}+\eta}
\prod_{n=1 \atop n\ne k}^{N_1}(z_{j}-u_{n}+\eta) 
\rr
\,, \non \\
{\cM}_{N_{2}+2j,k} &=& \frac{1 }{(u_{k}-z_{j}-\eta)}
\prod_{n=1 \atop n\ne k}^{N_1}(z_{j}-u_{n}) \,,  \qquad 
j = 1, \ldots, \frac{1}{2}(N_{1}-N_{2}) \,.
\ee 

\noindent We now observe that $\det {\cM}_{lk}$ does not change if 
we add to ${\cM}_{N_{2}+2j-1,k}$ any $k$-independent factor times 
${\cM}_{N_{2}+2j,k}$. The second term of ${\cM}_{N_{2}+2j-1,k}$ 
can therefore be dropped, since it can be written as

\be
-\frac{1 }{(u_{k}-z_{j})(z_{j}-u_{k}+\eta)}
\prod_{l=1}^{\LL_{1}}\frac{z_{j}-z_{l}-\eta}{z_{j}-z_{l}+\eta}
\prod_{n=1 \atop}^{N_1}(z_{j}-u_{n}+\eta) \,, \non
\ee 

\noindent which is a $k$-independent factor times 
${\cM}_{N_{2}+2j,k}$. In short, for $l > N_{2}$, ${\cM}_{lk}$
is given by 

\be
{\cM}_{N_{2}+2j-1,k} &=&  \frac{1 }{(u_{k}-z_{j})}
\prod_{n=1 \atop n\ne k}^{N_1}(z_{j}-u_{n}-\eta) \,, 
\label{restSlavnov2b} \\
{\cM}_{N_{2}+2j,k} &=& \frac{1 }{(u_{k}-z_{j}-\eta)}
\prod_{n=1 \atop n\ne k}^{N_1}(z_{j}-u_{n}) \,, 
\label{restSlavnov2c}  \qquad  
j = 1, \ldots, \frac{1}{2}(N_{1}-N_{2}) \,. 
\ee 

\begin{figure}
\begin{centering}
\includegraphics[height=5cm]{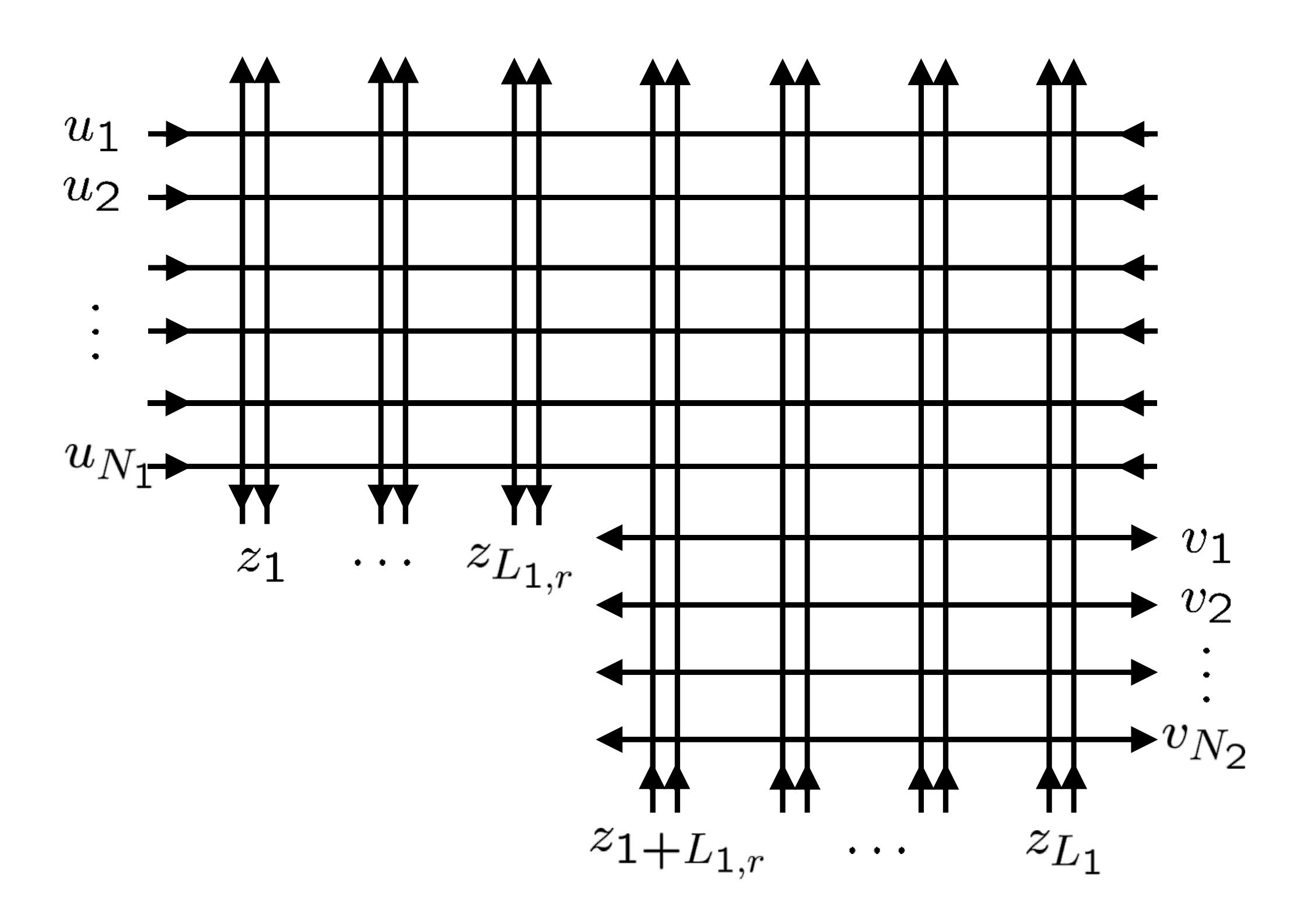}
\par\end{centering}
\caption{2D lattice representation of the restricted Slavnov
determinant, which is a scalar product between $|{\cO}_1\rangle$
(where the $B$ operators with arguments $\{u\}_{N_1}$ act on all 
the quantum spaces $1 \ldots, L_{1}$ ) and ${ }_r\langle {\cO}_2|$ 
(where the $C$ operators with arguments $\{v\}_{N_2}$ act only on 
the quantum spaces $1+L_{1,r}, 
\ldots, L_{1}$).}
\label{fig:fig4}
\end{figure}

\noindent With the help of the vertex-model correspondence, we can 
make the identification

\be
\tilde S_{restricted}=
\langle 1, \ldots, \frac{1}{2}(N_{1}-N_{2})| 
\prod_{j=1}^{N_{1}}\tilde C[v_{j}; \{z\}_{\LL_{1}}] 
\prod_{k=1}^{N_{1}}\tilde B[u_{k}; \{z\}_{\LL_{1}}]  |0 \rangle \,,
\ee 

\noindent where $| 1, \ldots, \frac{1}{2}(N_{1}-N_{2}) \rangle$ is 
the state with down-spins at the sites 
$1, \ldots, \frac{1}{2}(N_{1}-N_{2})$ 
and up-spins at the remaining $\LL_{1} - \frac{1}{2}(N_{1}-N_{2})$ 
sites. See Figure {\bf \ref{fig:fig4}}. Finally, returning to the 
original normalization using Equation ({\bf \ref{renBC}}), we obtain 

\begin{multline}
S[\{u\}_{N_1}, \{v\}_{N_2}, \{z\}_{\LL_1}] = 
\langle 1, \ldots, \frac{1}{2}(N_{1}-N_{2})| 
\prod_{j=1}^{N_{2}} C[v_{j}; \{z\}_{\LL_{1}}]  
\prod_{k=1}^{N_{1}} B[u_{k}; \{z\}_{\LL_{1}}] 
|0 \rangle
\\
= 
\ll
\prod_{l=\frac{1}{2}(N_{1}-N_{2})+1}^{\LL_1}
\prod_{k=1}^{N_{2}}
\frac{v_{k}-z_{l}+\eta}{v_{k}-z_{l}-\eta}
\rr
\prod_{N_{1}\ge j>k\ge 1}\frac{1}{u_{j}-u_{k}} 
\prod_{N_{2}\ge j>k\ge 1} \frac{1}{v_{j}-v_{k}} 
\\
\times \prod_{\frac{1}{2}(N_{1}-N_{2})\ge j>k\ge 1} 
\frac{1}{(z_{j}-z_{k})^{2}(z_{j}-z_{k}-\eta)(z_{j}-z_{k}+\eta)} 
\\
\times 
\prod_{j=1}^{\frac{1}{2}(N_{1}-N_{2})}
\prod_{k=1}^{N_{2}}\frac{1}{(z_{j}-v_{k}+\eta)(z_{j}-v_{k})} 
\det {\cM}_{lk} \,.
\label{restSlavnov}
\end{multline}

\begin{thebibliography}{10}

\bibitem{Minahan:2002ve}
J.~A. Minahan and K.~Zarembo, ``{The Bethe-ansatz for $\mathcal{N}$ = 4 super
  Yang-Mills},'' {\em JHEP} {\bfseries 03} (2003) 013,
\href{http://arxiv.org/abs/hep-th/0212208}{{\ttfamily arXiv:hep-th/0212208}}.

\bibitem{Beisert:2003yb}
N.~Beisert and M.~Staudacher, ``{The $\mathcal{N}$ = 4 SYM Integrable Super
  Spin Chain},'' \href{http://dx.doi.org/10.1016/j.nuclphysb.2003.08.015}{{\em
  Nucl. Phys.} {\bfseries B670} (2003) 439--463},
\href{http://arxiv.org/abs/hep-th/0307042}{{\ttfamily arXiv:hep-th/0307042}}.

\bibitem{Beisert:2005fw}
N.~Beisert and M.~Staudacher, ``{Long-range PSU(2,2$/$4) Bethe ansaetze for
  gauge theory and strings},''
  \href{http://dx.doi.org/10.1016/j.nuclphysb.2005.06.038}{{\em Nucl. Phys.}
  {\bfseries B727} (2005) 1--62},
\href{http://arxiv.org/abs/hep-th/0504190}{{\ttfamily arXiv:hep-th/0504190}}.

\bibitem{Beisert:2010jr}
N.~Beisert, C.~Ahn, L.~F. Alday, Z.~Bajnok, J.~M. Drummond, {\em et al.},
  ``{Review of AdS/CFT Integrability: An Overview},''
  \href{http://dx.doi.org/10.1007/s11005-011-0529-2}{{\em Lett.Math.Phys.}
  {\bfseries 99} (2012) 3--32},
\href{http://arxiv.org/abs/1012.3982}{{\ttfamily arXiv:1012.3982 [hep-th]}}.

\bibitem{Ferretti:2004ba}
G.~Ferretti, R.~Heise, and K.~Zarembo, ``{New integrable structures in large-N
  QCD},'' \href{http://dx.doi.org/10.1103/PhysRevD.70.074024}{{\em Phys. Rev.}
  {\bfseries D70} (2004) 074024},
\href{http://arxiv.org/abs/hep-th/0404187}{{\ttfamily arXiv:hep-th/0404187}}.

\bibitem{Zamolodchikov:1980ku}
A.~Zamolodchikov and V.~Fateev, ``{Model factorized S matrix and an integrable
  Heisenberg chain with spin 1. (in russian)},''
{\em Sov.J.Nucl.Phys.} {\bfseries 32} (1980) 298--303.

\bibitem{Kulish:1981gi}
P.~Kulish, N.~Reshetikhin, and E.~Sklyanin, ``{Yang-Baxter Equation and
  Representation Theory. 1.},''
  \href{http://dx.doi.org/10.1007/BF02285311}{{\em Lett.Math.Phys.} {\bfseries
  5} (1981) 393--403}.

\bibitem{Kulish:1981bi}
P.~Kulish and E.~Sklyanin, ``{Quantum spectral transform method. Recent
  developments},''
{\em Lect.Notes Phys.} {\bfseries 151} (1982) 61--119.

\bibitem{A.:1982zz}
L.~Takhtajan, ``{The picture of low-lying excitations in the isotropic
  Heisenberg chain of arbitrary spins},''
{\em Phys.Lett.} {\bfseries A87} (1982) 479--482.

\bibitem{Babujian:1983ae}
H.~M. Babujian, ``{Exact solution of the isotropic Heisenberg chain with
  arbitrary spins: thermodynamics of the model},''
\href{http://dx.doi.org/10.1016/0550-3213(83)90668-5}{{\em Nucl.Phys.}
  {\bfseries B215} (1983) 317--336}.

\bibitem{Lee:1998bxa}
S.~Lee, S.~Minwalla, M.~Rangamani, and N.~Seiberg, ``{Three point functions of
  chiral operators in D = 4, N=4 SYM at large N},'' {\em Adv.Theor.Math.Phys.}
  {\bfseries 2} (1998) 697--718,
\href{http://arxiv.org/abs/hep-th/9806074}{{\ttfamily arXiv:hep-th/9806074
  [hep-th]}}.

\bibitem{Okuyama:2004bd}
K.~Okuyama and L.-S. Tseng, ``{Three-point functions in $\mathcal{N}$ = 4 SYM
  theory at one-loop},''
  \href{http://dx.doi.org/10.1088/1126-6708/2004/08/055}{{\em JHEP} {\bfseries
  08} (2004) 055},
\href{http://arxiv.org/abs/hep-th/0404190}{{\ttfamily arXiv:hep-th/0404190}}.

\bibitem{Roiban:2004va}
R.~Roiban and A.~Volovich, ``{Yang-Mills correlation functions from integrable
  spin chains},'' \href{http://dx.doi.org/10.1088/1126-6708/2004/09/032}{{\em
  JHEP} {\bfseries 09} (2004) 032},
\href{http://arxiv.org/abs/hep-th/0407140}{{\ttfamily arXiv:hep-th/0407140}}.

\bibitem{Alday:2005nd}
L.~F. Alday, J.~R. David, E.~Gava, and K.~S. Narain, ``{Structure constants of
  planar $\mathcal{N}$ = 4 Yang Mills at one loop},'' {\em JHEP} {\bfseries 09}
  (2005) 070,
\href{http://arxiv.org/abs/hep-th/0502186}{{\ttfamily arXiv:hep-th/0502186}}.

\bibitem{Escobedo:2010xs}
J.~Escobedo, N.~Gromov, A.~Sever, and P.~Vieira, ``{Tailoring Three-Point
  Functions and Integrability},''
  \href{http://dx.doi.org/10.1007/JHEP09(2011)028}{{\em JHEP} {\bfseries 1109}
  (2011) 028},
\href{http://arxiv.org/abs/1012.2475}{{\ttfamily arXiv:1012.2475 [hep-th]}}.

\bibitem{Foda:2011rr}
O.~Foda, ``{N=4 SYM structure constants as determinants},''
  \href{http://dx.doi.org/10.1007/JHEP03(2012)096}{{\em JHEP} {\bfseries 1203}
  (2012) 096},
\href{http://arxiv.org/abs/1111.4663}{{\ttfamily arXiv:1111.4663 [math-ph]}}.

\bibitem{Beisert:2004fv}
N.~Beisert, G.~Ferretti, R.~Heise, and K.~Zarembo, ``{One-loop QCD spin chain
  and its spectrum},''
  \href{http://dx.doi.org/10.1016/j.nuclphysb.2005.04.004}{{\em Nucl. Phys.}
  {\bfseries B717} (2005) 137--189},
\href{http://arxiv.org/abs/hep-th/0412029}{{\ttfamily arXiv:hep-th/0412029}}.

\bibitem{Braun:2003rp}
V.~M. Braun, G.~P. Korchemsky, and D.~Mueller, ``{The uses of conformal
  symmetry in QCD},''
  \href{http://dx.doi.org/10.1016/S0146-6410(03)90004-4}{{\em Prog. Part. Nucl.
  Phys.} {\bfseries 51} (2003) 311--398},
\href{http://arxiv.org/abs/hep-ph/0306057}{{\ttfamily arXiv:hep-ph/0306057}}.

\bibitem{Jaffe:1991ra}
R.~Jaffe and X.-D. Ji, ``{Chiral odd parton distributions and Drell-Yan
  processes},''
\href{http://dx.doi.org/10.1016/0550-3213(92)90110-W}{{\em Nucl.Phys.}
  {\bfseries B375} (1992) 527--560}.

\bibitem{Braun:1998id}
V.~M. Braun, S.~E. Derkachov, and A.~Manashov, ``{Integrability of three
  particle evolution equations in QCD},''
  \href{http://dx.doi.org/10.1103/PhysRevLett.81.2020}{{\em Phys.Rev.Lett.}
  {\bfseries 81} (1998) 2020--2023},
\href{http://arxiv.org/abs/hep-ph/9805225}{{\ttfamily arXiv:hep-ph/9805225
  [hep-ph]}}.

\bibitem{Braun:1999te}
V.~M. Braun, S.~E. Derkachov, G.~Korchemsky, and A.~Manashov, ``{Baryon
  distribution amplitudes in QCD},''
  \href{http://dx.doi.org/10.1016/S0550-3213(99)00265-5}{{\em Nucl.Phys.}
  {\bfseries B553} (1999) 355--426},
\href{http://arxiv.org/abs/hep-ph/9902375}{{\ttfamily arXiv:hep-ph/9902375
  [hep-ph]}}.

\bibitem{Belitsky:1999qh}
A.~V. Belitsky, ``{Fine structure of spectrum of twist - three operators in
  QCD},'' \href{http://dx.doi.org/10.1016/S0370-2693(99)00326-3}{{\em
  Phys.Lett.} {\bfseries B453} (1999) 59--72},
\href{http://arxiv.org/abs/hep-ph/9902361}{{\ttfamily arXiv:hep-ph/9902361
  [hep-ph]}}.

\bibitem{Ahn:2007ky}
C.~Ahn, R.~I. Nepomechie, and J.~Suzuki, ``{The QCD spin chain S matrix},''
  \href{http://dx.doi.org/10.1016/j.nuclphysb.2007.12.026}{{\em Nucl. Phys.}
  {\bfseries B798} (2008) 402--422},
\href{http://arxiv.org/abs/0711.2415}{{\ttfamily arXiv:0711.2415 [hep-th]}}.

\bibitem{Belitsky:2005bu}
A.~V. Belitsky, G.~P. Korchemsky, and D.~Mueller, ``{Integrability of two-loop
  dilatation operator in gauge theories},''
  \href{http://dx.doi.org/10.1016/j.nuclphysb.2005.11.015}{{\em Nucl. Phys.}
  {\bfseries B735} (2006) 17--83},
\href{http://arxiv.org/abs/hep-th/0509121}{{\ttfamily arXiv:hep-th/0509121}}.

\bibitem{Gromov:2012vu}
N.~Gromov and P.~Vieira, ``{Quantum Integrability for Three-Point Functions},''
\href{http://arxiv.org/abs/1202.4103}{{\ttfamily arXiv:1202.4103 [hep-th]}}.

\bibitem{LimaSantos:1998te}
A.~Lima-Santos, ``{Bethe ansatz for nineteen vertex models},''
  \href{http://dx.doi.org/10.1088/0305-4470/32/10/004}{{\em J.Phys.A}
  {\bfseries A32} (1999) 1819--1839},
\href{http://arxiv.org/abs/hep-th/9807219}{{\ttfamily arXiv:hep-th/9807219
  [hep-th]}}.

\bibitem{Crampe:2010xx}
N.~Cramp\'e, E.~Ragoucy, and L.~Alonzi, ``{Coordinate Bethe Ansatz for Spin s
  XXX Model},'' {\em Sigma} {\bfseries 7} (2011) 006,
  \href{http://arxiv.org/abs/1009.0408}{{\ttfamily arXiv:1009.0408 [math-ph]}}.

\bibitem{Ovchinnikov:2010vb}
A.~Ovchinnikov, ``{Coordinate space wave function from the Algebraic Bethe
  Ansatz for the inhomogeneous six-vertex model},''
  \href{http://dx.doi.org/10.1016/j.physleta.2010.01.022}{{\em Phys.Lett.}
  {\bfseries A374} (2010) 1311--1314},
\href{http://arxiv.org/abs/1001.2672}{{\ttfamily arXiv:1001.2672 [math-ph]}}.

\bibitem{Izergin:1981aa}
A.~G. Izergin, ``{Partition function of the six-vertex model in a finite
  volume},'' {\em Sov. Phys. Dokl.} {\bfseries 32} (1987) 878--879.

\bibitem{Korepin:1982gg}
V.~E. Korepin, ``{Calculation of norms of bethe wave functions},''
\href{http://dx.doi.org/10.1007/BF01212176}{{\em Commun. Math. Phys.}
  {\bfseries 86} (1982) 391--418}.

\bibitem{Korepin:1993aa}
V.~Korepin, N.~Bogoliubov, and A.~Izergin, {\em {Quantum inverse scattering
  method and correlation functions}}.
\newblock Cambridge University Press, 1993.

\bibitem{Slavnov:1989aa}
N.~Slavnov, ``{Calculation of scalar products of wave functions and form
  factors in the framework of the algebraic Bethe ansatz},'' {\em Theor. Math.
  Phys.} {\bfseries 79} (1989) 502--508.

\bibitem{Gaudin:1976aa}
M.~Gaudin, ``{Diagonalisation of a class of spin hamiltonians},'' {\em Journal
  de Physique} {\bfseries 37} (1976) 1087--1098.

\bibitem{Gaudin:1981aa}
M.~Gaudin, B.~M. McCoy, and T.~T. Wu, ``{Normalization sum for the Bethe's
  hypothesis wave functions of the Heisenberg-Ising chain},'' {\em Phys. Rev.
  D} {\bfseries 23} (1981) 417--419.

\bibitem{Kitanine:1999aa}
N.~Kitanine, J.~Maillet, and V.~Terras, ``{Form factors of the XXZ Heisenberg
  spin-1/2 finite chain},'' {\em Nucl.Phys.} {\bfseries B554} (1999) 647--678,
  \href{http://arxiv.org/abs/9807020}{{\ttfamily arXiv:9807020 [math-ph]}}.

\bibitem{Wheeler:2011xw}
M.~Wheeler, ``{An Izergin-Korepin procedure for calculating scalar products in
  six-vertex models},''
  \href{http://dx.doi.org/10.1016/j.nuclphysb.2011.07.006}{{\em Nucl.Phys.}
  {\bfseries B852} (2011) 468--507},
\href{http://arxiv.org/abs/1104.2113}{{\ttfamily arXiv:1104.2113 [math-ph]}}.

\end{thebibliography}

\providecommand{\href}[2]{#2}\begingroup\raggedright\endgroup

\end{document}